\pdfoutput=1
\documentclass[letterpaper,twocolumn,10pt]{article}
\usepackage{usenix}

\usepackage{tikz}
\usepackage{amsmath}
\usepackage{amssymb}
\usepackage{enumitem}
\usepackage{subfiles}
\usepackage{booktabs}
\usepackage[flushleft]{threeparttable}
\usepackage[font=bf]{caption}
\usepackage{filecontents}
\usepackage{ifthen}
\usepackage{xspace}
\usepackage{titling}

\usepackage[compact]{titlesec}
\titlespacing{\section}{0pt}{2ex}{1ex}
\titlespacing{\subsection}{0pt}{0.5ex}{0.5ex}
\titlespacing{\subsubsection}{0pt}{1ex}{1ex}
\usepackage{titlesec}
\titleformat{\section}{\normalfont\fontsize{12}{15}\bfseries}{\thesection}{1em}{}
\titleformat{\subsection}{\normalfont\fontsize{11}{15}\bfseries}{\thesubsection}{0.5em}{}

\usepackage{enumitem}
\setlist{itemsep=0pt,parsep=0pt}             

\usepackage{fancyhdr}
\newcommand{\system}{\textsc{Amadeus}\xspace}

\newcommand{\PAGENUMBERS}{yes}       

\newenvironment{smitemize}%
  {\begin{list}{$\bullet$}%
     {\setlength{\parsep}{0pt}%
      \setlength{\topsep}{0pt}%
      \setlength{\itemsep}{0pt}}}%
  {\end{list}}

\ifthenelse{\equal{\PAGENUMBERS}{yes}}{%
	\pagestyle{plain}
}{%
\pagestyle{empty}
}

\date{}
\ifthenelse{\equal{\PAGENUMBERS}{yes}}{%
	\thispagestyle{fancy}
}{%
\thispagestyle{empty}
}

\title{\bf \Large \system: Scalable, Privacy-Preserving Live Video Analytics} 

\author{
{\rm Sandeep D'souza}\\
Carnegie Mellon University
\and
{\rm Victor Bahl}\\
Microsoft Research
\and
{\rm Lixiang Ao}\\
UC San Diego
\and
{\rm Landon P. Cox}\\
Microsoft Research
} 
\begin{document}
\maketitle

\subsection*{Abstract}
Smart-city applications ranging from traffic management to public-safety alerts rely on live analytics of video from surveillance cameras in public spaces. However, a growing number of government regulations stipulate how data collected from these cameras must be handled in order to protect citizens' privacy. This paper describes \system, which balances privacy and utility by redacting video in near realtime for smart-city applications. Our main insight is that whitelisting objects, or blocking by default, is crucial for scalable, privacy-preserving video analytics. In the context of modern object detectors, we prove that whitelisting \textit{reduces} the risk of an object-detection error leading to a privacy violation, and helps \system scale to a large and diverse set of applications. In particular, \system utilizes whitelisting to generate \textit{composable} encrypted object-specific live streams, which simultaneously meet the requirements of multiple applications in a privacy-preserving fashion, while reducing the compute and streaming-bandwidth requirements at the edge. Experiments with our \system prototype show that compared to blacklisting objects, whitelisting yields significantly better privacy (up to \textasciitilde28x) and bandwidth savings (up to \textasciitilde5.5x). Additionally, our experiments also indicate that the composable live streams generated by \system are usable by real-world applications with minimum utility loss.

\section{Introduction} \label{sec:intro}
Surveillance cameras are a ubiquitous presence in public spaces. To take advantage of these cameras, researchers have proposed several video-analytics frameworks~\cite{rocket, focus, noscope} that can ingest video data and perform object detection~\cite{yolo, faster-rcnn} and tracking~\cite{sort, opencv-trackers-1} on behalf of smart-city applications like traffic management~\cite{quartz}, pedestrian detection, and public-safety alerts. Cameras are an appealing data source for smart-city initiatives because they can support a wide-range of applications and installing cameras is relatively easy. In comparison, dedicated sensing infrastructure, such as in-road induction loops or RFID tags and readers, support a far smaller set of use-cases and are more difficult to deploy.

However, as we have learned from our own experience building pilot video-analytics systems for the last several years, the convenience of surveillance cameras comes at a significant cost: erosion of citizens' privacy. Cameras collect data indiscriminately, and as a result they capture information that is both inessential to applications' purposes and prone to abuse. 
An increasing cause of worry is applications going beyond their mandate or publicly-specified objective. Consider a smart-city scenario, where a video stream originating at a camera may be processed by third-party entities, each with their own mandate. For example, consumer \textsf{A} is mandated to analyze vehicles to measure traffic congestion, and consumer \textsf{B} is mandated to count pedestrians to decide an appropriate pedestrian-crossing duration. 
However, consumer \textsf{A} may exceed their mandate by performing facial recognition on the video stream, without the administrator's knowledge. 

Governments around the world have reacted to the privacy threat posed by surveillance technologies by passing regulations like the European Union's General Data Protection Regulations (GDPR) and cities' bans on face recognition~\cite{sf-face-recognition, somerville-face-recognition}. Preserving the utility of video-based smart-city applications while remaining compliant with privacy-protecting regulations will require technical solutions for controlling how applications extract information from video data.

Smartphones and PCs must also control access to video data, but they typically support only coarse-grained permissions, e.g., allowing full access to a device's camera or no access at all. Fortunately, prior work has explored applying the principle of least privilege to video using fine-grained access-control mechanisms. These systems rely on computer vision to detect objects in realtime or near realtime~\cite{recognizer, surround-web, i-pic, private-eye, scanner-darkly, life-log}, and then use the detected objects to transform raw video data before sharing it with an application. One approach to transforming video data is \emph{blacklisting} or \emph{sharing-by-default}, which redacts objects that a video consumer is not allowed to view (e.g., placing a black box over human faces)~\cite{sensorsift, life-log, i-pic}. Another approach is \emph{whitelisting} or \emph{blocking-by-default}, which shares only the objects that a video consumer is explicitly allowed to view (e.g., blacking out everything except for detected cars and trucks)~\cite{recognizer, private-eye, scanner-darkly, surround-web}.

These techniques provide a solid foundation for bringing smart-city infrastructure into compliance with privacy regulations, but prior work is insufficient on its own. In particular, prior work was designed for smartphones and PCs, in which a trusted kernel can transform video data for a small number of local applications (often just one). In contrast, surveillance-camera data circulates through a distributed system consisting of resource-limited edge devices and a potentially diverse and large set of remote video-consuming applications. 

In this paper we present the design, implementation, and evaluation of a privacy-preserving framework for surveillance-camera videos called \system. The three biggest challenges faced by \system are edge devices' compute limitations, edge devices' bandwidth limitations, and smart-cities' large and diverse set of video consumers. The key observation underlying our design is that a block-by-default approach is crucial for preserving privacy in the face of these challenges. 

To limit the size of \system's trusted computing base, \system redacts videos on the edge using off-the-shelf object-detection models. However, performing object detection on resource-limited edge devices in near realtime often means using \textit{weaker} models. Our experiments (Section \ref{sec:implications-privacy}) indicate that weaker models are more prone to false negatives (i.e., missing a present object) than false positives (i.e., misclassifying a present object), and these errors can lead to privacy violations. Thus, \system blocks video content by default, or \textit{whitelists} objects, to reduce the likelihood of a model error causing a privacy violation.

Blocking by default (whitelisting) also helps \system scale to a large and diverse set of applications. Modern object-detection models like YOLO~\cite{yolo} can detect tens of object categories, such as people, cars, and traffic lights. Under share-by-default (blacklisting), \system would have to create a uniquely-redacted stream for each combination of categories. Creating all of these streams would require a prohibitive amount of computation and bandwidth for an edge node. Note that even when the number of permission combinations for active near realtime applications is small, the diversity of authorized video consumers may grow over time. For example, at the time a camera captures a video, no active application may need to view bicycles. However, at a later time, a city's road-planning commission may wish to characterize bicycle traffic using archived videos.

As a result, \system creates a separate live stream for each object category (along with a residual background stream), and each stream contains nothing but its associated category (e.g., the bicycle stream only shows bicycles). This approach helps \system scale in three ways. First, the maximum number of streams that an edge server needs to generate is equal to the number of detected object categories (not the number of category combinations). Second, because most streams are mostly (or entirely) blacked out, individual streams compress well with minimum effort. And finally, by breaking object streams into short segments and encrypting the segments with unique keys, \system can decouple serving encrypted video data from making authorization decisions (i.e., distributing decryption keys). This allows video transfers to be handled by conventional, scalable web technologies like content distribution networks (CDNs) without expanding the trusted computing base or undermining privacy. Applications authorized to view multiple objects can simply retrieve the appropriate keys and encrypted video segments, and overlay the decrypted videos to compose a coherent, multi-object stream. 

\system also provides object-specific encrypted live analytics which enable applications to extract insights without re-processing the video streams. This metadata also indicates the presence of low-confidence relevant objects withheld from the consumer, which can be used to recover lost utility. 

\textbf{Threat Model:} We assume that applications are trusted to perform \textit{any} operations on object streams they are \textit{authorized} to view. For instance, an application authorized to view an object stream corresponding to the class \textsf{faces}, can perform any operation on \textsf{faces} (such as facial recognition and age/gender/emotion detection). Therefore, \system authorizes applications' purpose at the granularity of the detection classes supported by the object detector. Providing more fine-grained authorization requires feature extraction \cite{Ossia2020DeepPE}, which reduces flexibility, and is beyond the scope of this work.   

This paper makes the following contributions:
\begin{smitemize}
  \item We identify block-by-default or whitelisting as crucial for scalable, privacy-preserving live video analytics. 
  \item We prove that whitelisting can \textit{guarantee} lower privacy leakage than blacklisting, in the context of modern object-detection techniques.
  \item We describe the design of a new privacy-preserving video-analytics framework called \system that transforms video in near realtime and decouples video authorization from video transfer, using \textit{composable} streams.
  \item Using our \system prototype we show that blocking-by-default balances privacy and utility and significantly reduces the bandwidth required to simultaneously serve multiple applications with different objectives.
\end{smitemize}

\section{Privacy and Object Detection} \label{sec:background}

We now illustrate the benefits of whitelisting over blacklisting in the context of preserving privacy, by experimentally analyzing the performance of state-of-the-art object detectors.

Object detection is key to generating whitelisted and blacklisted video streams. Therefore, the object detector utilized at video-ingest time plays an important role in deciding:
\setlist[enumerate]{noitemsep, nolistsep}
\begin{enumerate}[noitemsep]
	\item the amount of private information that will be \textit{disclosed} to the consumer, or \textit{privacy loss}, due to \textit{private} objects not being redacted (blacklisting), or being accidentally disclosed (whitelisting); and
	\item the amount of useful information that will be \textit{witheld} from the consumer, or \textit{utility loss}, due to \textit{relevant} objects not being disclosed (whitelisting), or being accidentally redacted (blacklisting)
\end{enumerate}

Convolutional Neural Networks (CNNs) \cite{cnn} are the basis for most state-of-the-art techniques for both object-detection \cite{yolo, rcnn, ssd} and image-classification \cite{resnet} tasks. 


Object-detection techniques in the family of Region-based CNNs (RCNNs) \cite{rcnn, fast-rcnn, faster-rcnn} utilize a two-stage approach. In the first stage, the model proposes a set of regions of interest. Subsequently, an image classifier assigns a label to each of these regions. On the other hand, one-stage detectors like SSD \cite{ssd} and YOLO \cite{yolo, yolo-v2, yolo-v3} omit the region-proposal stage and run detection directly over a dense sampling of possible locations over the image. These one-stage detectors lead to faster detection speeds at the cost of slightly lower accuracy. 

In this paper we consider Faster-RCNN \cite{faster-rcnn} (a two-stage detector) and YOLO \cite{yolo-v3} (a one-stage detector), as most modern object detectors are modified versions of these models.

\subsection{Implications on Privacy and Utility} \label{sec:implications-privacy}
We now look at the implications of object-detection accuracy on privacy and utility. Every detection/mis-detection of an object detector can be categorized as follows:
\setlist[itemize]{noitemsep, nolistsep}
\begin{itemize}[noitemsep]
	\item \textbf{True Positives (TP)}: the detector correctly detects and labels an object.
	\item \textbf{False Positives (FP)}: the detector mislabels an object.
	\item \textbf{False Negatives (FN)}: the detector misses an object and is unable to detect it.
\end{itemize}

\begin{figure}[t]
	\centering
	\includegraphics[width=0.6\columnwidth]{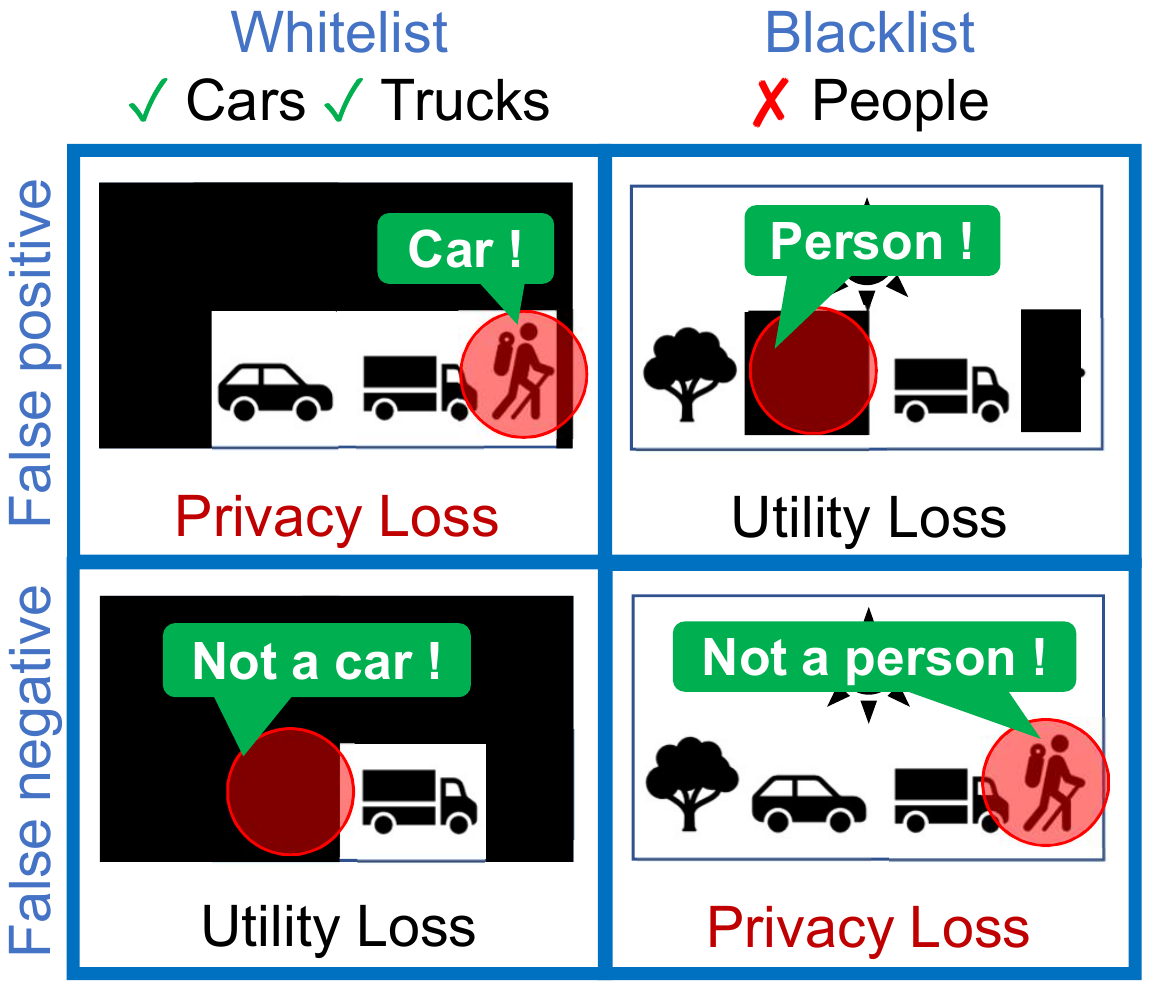} 
	\vspace{-0.2cm}
	\caption{The impact of false positives and false negatives} 
	\label{fig:confusion-matrix}
\end{figure}

Of the above three categories, false positives and false negatives constitute detection errors, and hence carry the potential to cause a loss in privacy or utility. Consider an application mandated to count vehicles. To minimize privacy loss, while ensuring that the application can still count vehicles: 
\setlist[itemize]{noitemsep, nolistsep}
\begin{itemize}[noitemsep]
	\item the \textit{whitelisting} approach only discloses objects detected as belonging to a class $R$ in the \textit{relevant} or whitelisted set $\omega$. In this case,  $\{$\textsf{car}, \textsf{truck}$\}\in\omega$.
	\item the \textit{blacklisting} approach redacts all objects detected as belonging to a class $S$ in the \textit{sensitive} or blacklisted set $\beta$.  In this case,  $\{$\textsf{person}$\}\in\beta$.
\end{itemize}

Figure \ref{fig:confusion-matrix} highlights the impact of false positives and false negatives on privacy and utility, using a vehicle-counting example. Consider whitelisting, where detecting a \textsf{person} and mis-classifying her as a \textsf{car} constitutes a false positive, leading to privacy loss for the person. On the other hand, not detecting a \textsf{car} in the scene constitutes a false negative, which leads to utility loss for the vehicle-counting application. For blacklisting, detecting a \textsf{car} as a \textsf{person} constitutes a false positive, which leads to utility loss for the vehicle-counting application. Alternatively, not detecting a \textsf{person} in the scene constitutes a false negative, which leads to privacy loss. 

The relative prevalence of false positives and false negatives decides whether whitelisting or blacklisting can lead to lower privacy loss. 
This is usually measured using precision and recall, which are defined as follows:
\setlist[itemize]{noitemsep, nolistsep}
\begin{itemize}[noitemsep]
	\item \textbf{Precision} measures how accurate a detector is, i.e., the proportion of detections that are correct. Precision can be calculated by: $precision=TP/(TP+FP)$. 
	\item \textbf{Recall} measures the fraction of ground truth objects that are both correctly detected and labeled by the detector. Recall can be calculate by: $recall = TP/(TP+FN)$.
\end{itemize}

Increasing precision involves decreasing the number of false positives (FP), and to increase recall the number of false negatives (FN) need to decrease. Note that most object detectors are trained to maximize mean-average precision \cite{ms-coco}.

\begin{figure}[t]
	\centering
	\includegraphics[width=\columnwidth]{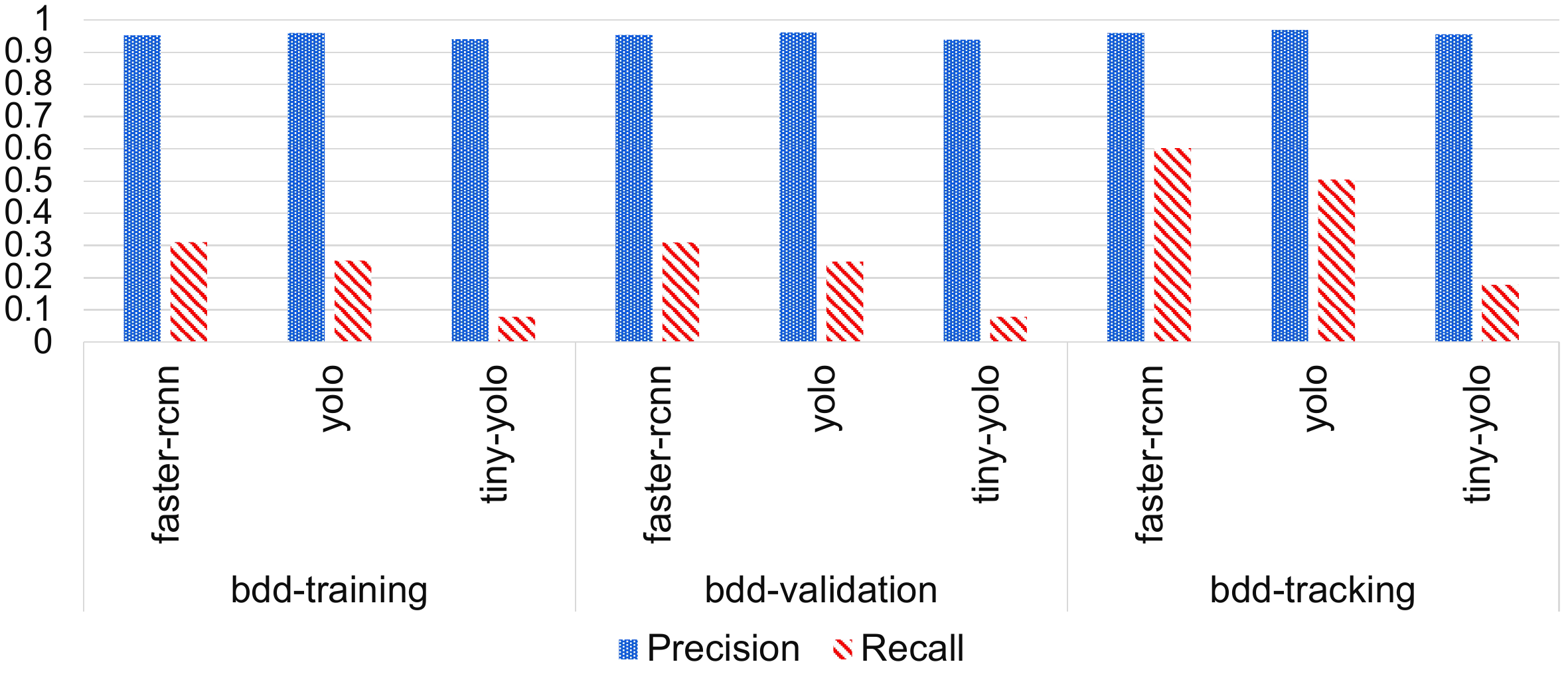} 
	\vspace{-0.6cm}
	\caption{BDD Dataset \cite{bdd}: Precision and Recall }
	\label{fig:precision-recall}
\end{figure}

Figure \ref{fig:precision-recall} plots the precision and recall of three object-detection techniques: Faster-RCNN \cite{faster-rcnn}, YOLOv3 and Tiny-YOLOv3 \cite{yolo-v3}, on three videos generated from the training (\textsf{bdd-training}), validation (\textsf{bdd-validation}), and tracking (\textsf{bdd-tracking}) data from the Berkeley DeepDrive Dataset (BDD) \cite{bdd}. These videos collectively contain 1,100 hours of labeled video data, consisting of 100,000 video sequences recorded at different times of the day, weather conditions, and scenarios \cite{bdd}. Thus, this dataset is representative of many real-world conditions and contains labeled objects corresponding to 10 classes (a subset of the COCO dataset \cite{ms-coco}). 

Among the object-detection techniques we benchmark, Faster-RCNN and YOLOv3 are relatively-heavy full-scale models, while Tiny-YOLOv3 is a lightweight approximation of the full-scale YOLOv3 model. We do not re-train these three models and use their default weights obtained by training on the COCO dataset \cite{ms-coco}, which has 80 object classes. This gives us an un-biased measurement of their performance on unseen data. We observe that, while all three detectors yield very high precision (\textasciitilde$0.93-0.96$), the recall is significantly lower (\textasciitilde$0.07-0.6$). For all three videos also observe that the most-complex model, i.e., Faster-RCNN yields the highest recall, followed by YOLOv3 and Tiny-YOLOv3.

\textbf{Observation 1:} Figure \ref{fig:precision-recall} indicates that state-of-the-art object detectors tend to have significantly higher precision than recall, i.e., false negatives are significantly more common than false positives for state-of-the-art object-detection techniques. This implies that object detectors tend to \textit{miss} more objects in a scene than \textit{mis-classify} detected objects. As indicated by Figure \ref{fig:precision-recall}, this problem is more acute for low-cost detectors like Tiny-YOLOv3, which have significantly lower recall (\textasciitilde 3x), and are often used in resource-scarce settings. 

We hypothesize that decreasing false positives is an easier problem to solve, as it involves \textit{correctly} classifying detected objects. This is supported by the results in Figure \ref{fig:precision-recall}, which indicate that light-weight detectors like Tiny-YOLOv3 have similar precision to computationally-heavy detectors like YOLOv3 and Faster-RCNN. On the other hand reducing false negatives is tougher, as it involves both detecting the \textit{missed} objects as well as classifying them. 

\subsection{Whitelisting: Privacy-Loss Guarantees} \label{sec:privacy-loss-guarantees}
While precision and recall are useful metrics to capture the prevalence of detection errors, as indicated in Figure \ref{fig:confusion-matrix}, not all such errors contribute to either privacy or utility loss.

\textbf{Observation 2:} A false positive $FP_{S\rightarrow R}$ mis-labeling a \textit{sensitive} class $S\in\beta$, as a \textit{relevant} class $R\in\omega$ causes privacy loss for whitelisting. Similarly, a false positive $FP_{R\rightarrow S}$ mis-labeling a \textit{relevant} class $R\in\omega$, as a \textit{sensitive} class $S\in\beta$ causes utility loss for blacklisting. On the other hand, a false negative $FN_S$ not detecting a sensitive class $S\in\beta$ causes privacy loss for blacklisting. Alternatively, a false negative $FN_R$ not detecting a relevant class $R\in\omega$ causes utility loss for whitelisting. We formally define privacy and utility loss:
\setlist[itemize]{noitemsep, nolistsep}
\begin{itemize}[noitemsep]
	\item \textbf{Privacy Loss}, $P$ is the fraction of objects belonging to sensitive classes $S \in \beta$ \textit{disclosed} to the consumer.
	\item \textbf{Utility Loss}, $U$ is the fraction of objects belonging to relevant classes $R\in \omega$ \textit{witheld} from the consumer.
\end{itemize}

Using the above definitions, we can mathematically calculate privacy and utility loss, $P$ and $U$, for both whitelisting ($WL$) and blacklisting ($BL$), using the following equations:
\begingroup
\begin{equation} \label{eqn:privacy-utility-eqn}
\begin{split}
P_{WL} = \frac{\sum_{S \in \beta, R \in \omega} FP_{S\rightarrow R}}{\sum_{S \in \beta}TP_S+FN_S}, \ P_{BL}  = \frac{\sum_{S \in \beta} FN_{S}}{\sum_{S \in \beta} TP_S+FN_S} \\ 
U_{WL} =  \frac{\sum_{R \in \omega}FN_{R}}{\sum_{R \in \omega} TP_R+FN_R}, \ U_{BL} = \frac{\sum_{S \in \beta, R \in \omega} FP_{R\rightarrow S}}{ \sum_{R \in \omega} TP_R+FN_R} 
\end{split}
\end{equation}
\endgroup


Combining Observations 1 and 2 we can conclude that:
\setlist[itemize]{noitemsep, nolistsep}
\begin{enumerate}[noitemsep]
	\item False positives are rare as modern object detectors have high precision. When using whitelisting, only a subset of false positives mis-labeling a \textit{sensitive} class $S\in\beta$ as a \textit{relevant} class $R\in\omega$ cause privacy loss.
	\item False negatives are common as modern object detectors have relatively-low recall. Although, recall can be increased by training on specialized data, it is significantly more challenging than increasing precision. 
	 When using blacklisting, \textit{all} false negatives not detecting a sensitive class $S\in\beta$ cause privacy loss.
\end{enumerate}

We now state the \textbf{\textsf{Privacy-Loss}} guarantee theorem.

\textbf{Theorem 1:} Consider objects $\Theta$ belonging to sensitive classes $S \in \beta$, which if disclosed lead to privacy loss. Given an object-detection technique with higher precision than recall over all objects $\Theta$ belonging to classes $S \in \beta$, \textit{then} whitelisting can \textit{guarantee} lower privacy loss than blacklisting.

\textit{Proof:} Using the definitions of precision and recall from Section \ref{sec:implications-privacy}, we can formulate the following inequality:
\begingroup
\setlength\abovedisplayskip{1pt}
\setlength\belowdisplayskip{1pt}
\begin{align*} 
precision_{\beta} &> recall_{\beta} \\ 
\Rightarrow \frac{\sum_{S \in \beta} TP_{S}}{\sum_{S \in \beta}TP_S+FP_S}  &>  \frac{\sum_{S \in \beta} TP_{S}}{\sum_{S \in \beta}TP_S+FN_S} \\
\Rightarrow \sum_{S \in \beta}FN_S &> \sum_{S \in \beta}FP_S
\end{align*}
\endgroup
The above inequality implies that the number of false negatives, $\sum_{S \in \beta}FN_S$ exceeds the number of false positives, $\sum_{S \in \beta}FP_S$, over all objects belonging to classes $S \in \beta$. We can also reformulate the above inequality as the following:
\begingroup
\setlength\abovedisplayskip{2pt}
\setlength\belowdisplayskip{2pt}
\begin{align*}
\frac{\sum_{S \in \beta} FP_{S}}{\sum_{S \in \beta}TP_S+FN_S} &< \frac{\sum_{S \in \beta} FN_{S}}{\sum_{S \in \beta} TP_S+FN_S} 
\end{align*}
\endgroup
As the number of false positives mis-classifying a relevant class $R \in \omega$ as a sensitive class $S \in \beta$ is a subset of the total false positives over all objects belonging to classes $S \in \beta$, i.e., $\sum_{S \in \beta, R \in \omega} FP_{S\rightarrow R} \leq \sum_{S \in \beta} FP_{S}$, the following holds:
\begingroup
\setlength\abovedisplayskip{2pt}
\setlength\belowdisplayskip{2pt}
\begin{align*}
\frac{\sum_{S \in \beta, R \in \omega} FP_{S\rightarrow R}}{\sum_{S \in \beta}TP_S+FN_S} &< \frac{\sum_{S \in \beta} FN_{S}}{\sum_{S \in \beta} TP_S+FN_S} \Rightarrow P_{WL} < P_{BL}
\end{align*}
\endgroup
where, $P_{WL}$ and $P_{BL}$ is the privacy loss caused when using whitelisting and blacklisting respectively. This follows from the definitions of privacy loss in Equation \ref{eqn:privacy-utility-eqn}. \hfill $\blacksquare$

\begin{figure}[t]
	\centering
	\includegraphics[width=\columnwidth]{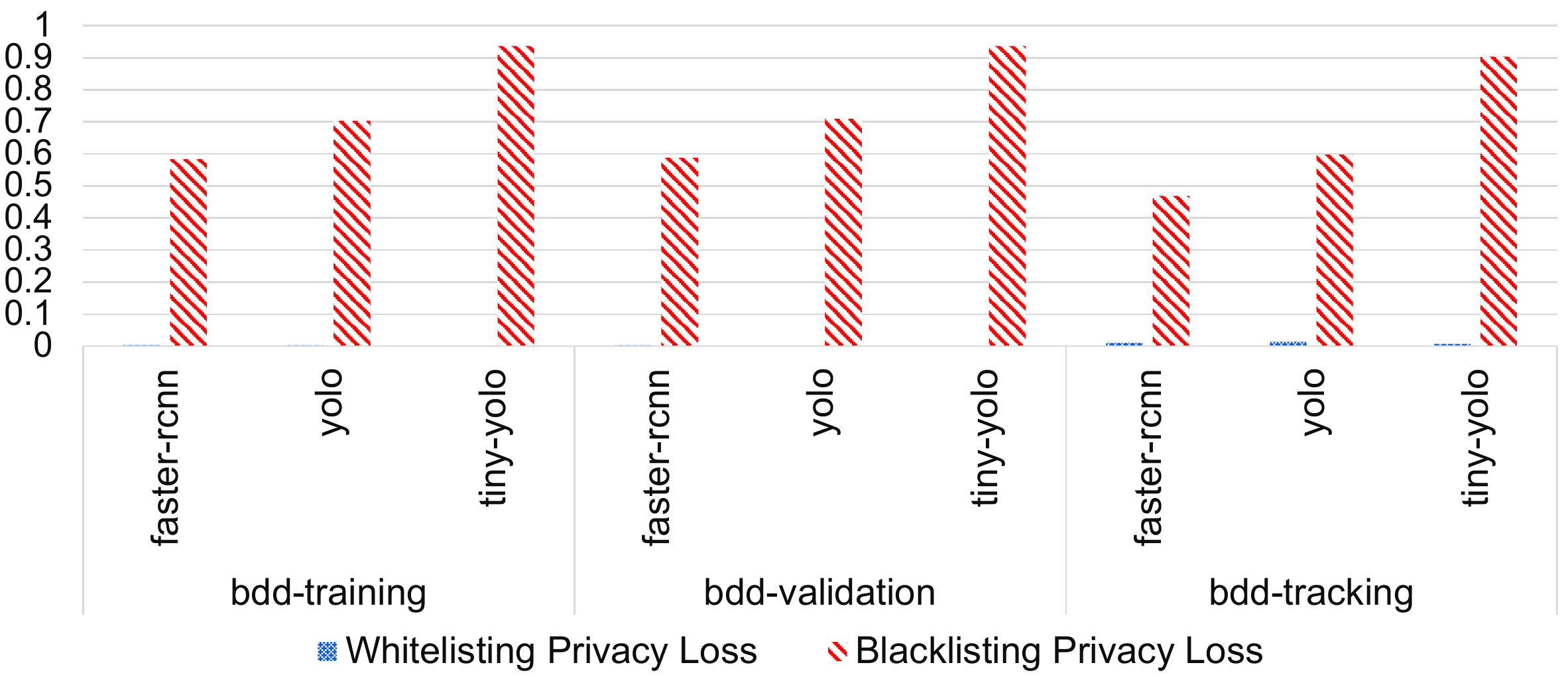} 
	\vspace{-0.6cm}
	\caption{Vehicle-Counting Privacy-Loss Measurements on the Berkeley DeepDrive Dataset \cite{bdd}}
	\label{fig:privacy-loss}
\end{figure}

\textbf{Key Takeaway:} Our observations coupled with the \textsf{Privacy-Loss} guarantee theorem indicate that when using modern object-detection methodologies at ingest time, whitelisting can yield significantly lower privacy loss than blacklisting. Figure \ref{fig:privacy-loss} compares the privacy loss between whitelisting and blacklisting, for the vehicle-counting application described in Figure \ref{fig:confusion-matrix}. 
We observe that whitelisting yields significantly lower percentage privacy loss (\textasciitilde$0.05$-$1.4\%$), as compared to blacklisting (\textasciitilde$46$-$93\%$). On the flip side, whitelisting also suffers potentially higher utility loss (\textasciitilde$38$-$87\%$) as compared to blacklisting (\textasciitilde$0.02$-$0.09\%$).

\textbf{Observation 3:} Privacy loss is permanent, utility loss can be recovered. For instance, when an object belonging to a sensitive class $S$ is accidentally disclosed or not redacted, the privacy loss caused cannot be reversed. On the other hand, if the detection of a low-confidence object belonging to a relevant class $R$ is withheld from the consumer, it can be recovered by post-processing the frame with a more-accurate object detector. This can be achieved by providing consumers with metadata indicating low-confidence relevant objects. Consumers can request post processing to recover lost utility.

Therefore, we bias towards privacy and utilize whitelisting over blacklisting for creating privacy-preserving video streams. Although our implementation supports both blacklisting and whitelisting, \system favors whitelisting. In subsequent sections, we highlight the added benefits whitelisting introduces in terms of both administering and distributing privacy-preserving video streams to multiple consumers.

\section{\system Design} \label{sec:design}
We now describe the motivation for \system's design, with a focus on the trade-offs of using \textit{whitelisting}. 
We assume that \system performs object detection using a surveillance camera's onboard computation~\cite{vision-ai} or a directly connected edge server~\cite{data-box-edge}. In either case, \system's trusted computing base (TCB) extends to all software and hardware with access to plaintext video. In the future, it may be possible to reduce the size of \system's TCB by performing object detection and video encoding in a secure execution environment, such as ARM TrustZone or Intel SGX, but this is currently infeasible. Regardless, because edge platforms are generally more resource constrained than high-end server machines, we explore some lightweight techniques to 
detect and disclose/redact objects in video streams. Subsequently, we describe how \system simultaneously delivers video to multiple applications, while minimizing privacy leakage and reducing the bandwidth required by the edge site. Lastly, we also discuss the benefits of whitelisting from the standpoint of application authorization.


\begin{figure}[t]
	\centering
	\includegraphics[width=\columnwidth]{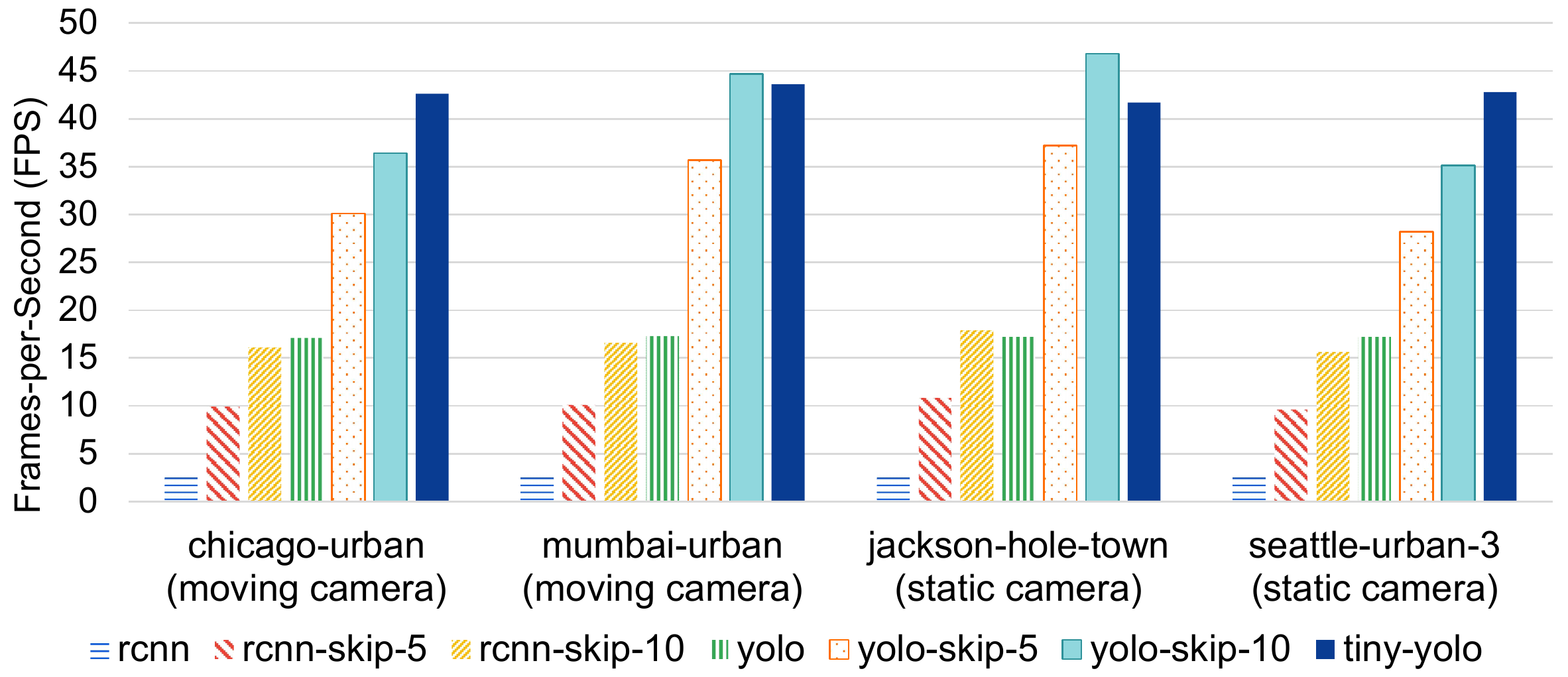} 
	\vspace{-0.6cm}
	\caption{Per-second frame-rate (FPS) for different object detection techniques combined with object tracking}
	\label{fig:tracking-fps}
\end{figure}

\subsection{Combining Detection with Tracking} \label{sec:tracking}
Most object detectors are computationally-intensive and require powerful hardware accelerators like GPUs to support inference frame rates required for real-time streaming and analytics. Such powerful accelerators may often not be available on resource-constrained edge platforms. A large body of work has looked at using neural-network specialization \cite{focus, nn-specialization-1, nn-specialization-2} and approximation \cite{nn-approx-1, nn-approx-2} to support fast inference.


This section explores utilizing existing pre-trained object-detection models coupled with low-cost object trackers. 
Given that objects remain in the field of view of a camera for \textit{atleast} a few frames, we perform inference every $n$ frames, and track the detected objects between two consecutive inference steps. 

In addition to tracking, other traditional techniques like frame differencing can also be used by \system to detect objects arriving between frames \cite{focus}. However, we focus on tracking as it also uniquely identifies object between frames, which can be utilized to service more sophisticated queries. 

Figure \ref{fig:tracking-fps} compares the inference frame rates of: (i) state-of-the-art object detectors: Faster-RCNN (\textsf{rcnn}) \cite{faster-rcnn} and YOLOv3 (\textsf{yolo}) \cite{yolo-v3}, (ii) a low-cost object detector: Tiny-YOLOv3 (\textsf{tiny-yolo}) \cite{yolo-v3}, and (iii) performing object detection every $n$ frames using  Faster-RCNN (\textsf{rcnn-skip-n}) or YOLOv3 (\textsf{yolo-skip-n}) followed by object tracking (using the \textsf{SORT} \cite{sort} framework and OpenCV-based trackers \cite{opencv-trackers}), for $n=5$ and $n=10$. These frame rates were obtained on a machine with an Nvidia GTX 1070 GPU \cite{gtx-1070}. We consider four videos each with an HD resolution of 1280x720. 

\begin{figure}[t]
	\centering
	\includegraphics[width=\columnwidth]{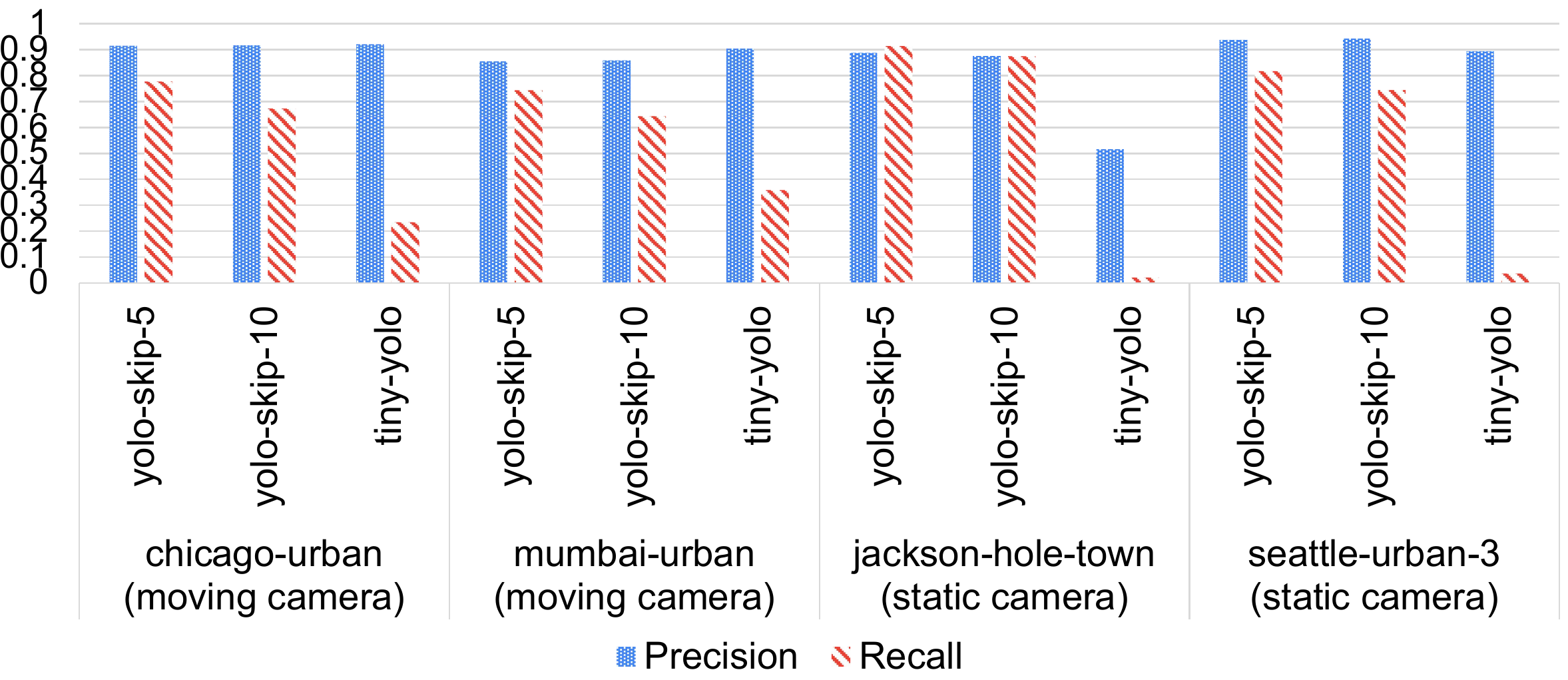} 
	\vspace{-0.6cm}
	\caption{Precision and Recall measurements for object detection (YOLOv3 \cite{yolo}) combined with object tracking}
	\label{fig:tracking-precision-recall}
\end{figure}

On average \textsf{rcnn} yields the lowest frame rate (\textasciitilde2.7 FPS), and \textsf{tiny-yolo} the fastest frame rate (\textasciitilde42 FPS). Additionally, \textsf{yolo-skip-10} can deliver a similar or faster frame rate than \textsf{tiny-yolo} (\textasciitilde36-46 FPS). Note that while \textsf{rcnn}, \textsf{yolo} and \textsf{tiny-yolo} yield frame rates independent of the video content, the frame rate for \textsf{rcnn-skip-n} and \textsf{yolo-skip-n} is video dependent, as the tracking speed depends on the number of objects in a frame. 

Figure \ref{fig:tracking-precision-recall} compares the precision and recall of \textsf{yolo-skip-n} for $n=5$ and $n=10$ and \textsf{tiny-yolo}. As the videos we consider are not labeled, we use labels generated by YOLOv3 (\textsf{yolo}) as the ground truth. This yields a good relative comparison, as all the techniques we compare are based on YOLOv3. We observe that \textsf{yolo-skip-n} for both $n=5$ and $n=10$, yields better precision and significantly higher recall as compared to \textsf{tiny-yolo}. This makes it a viable alternative for resource-constrained edge platforms, as it can deliver higher precision and recall than \textsf{tiny-yolo}, which translates to both lower privacy and utility loss, while delivering a similar or higher frame rate. 

\begin{table}[t]
	\centering
	\small
	\caption{Multi-consumer Smart-City Scenario}
	\vspace{-0.2cm}
	\label{tab:amadeus-scenario}
	\resizebox{\columnwidth}{!}{%
		\begin{threeparttable}
			\begin{tabular}{c|l|l}
				\toprule
				\textbf{Application} & \textbf{Whitelisted Objects} & \textbf{Blacklisted Objects}  \\ 
				\midrule
				Traffic Management   & vehicle & person  \\
				\midrule
				Safety Alerts        & person & vehicle  \\
				\midrule
				Bicycle Safety       & bicycle & vehicle, person  \\
				\midrule
				Two-wheeler Counting & bicycle, motorbike & vehicle, person  \\
				\midrule
				Surveillance         & all + background & none  \\
				
				\bottomrule	
			\end{tabular}
	\end{threeparttable}}
\end{table}

\subsection{Composable Streaming} \label{sec:amadeus}

Bandwidth is an important constraint when streaming from an edge device \cite{satya-edge, gigasight, shi2019edge}. If all consumers are homogeneous in terms of their viewing privileges and objectives, a single whitelisted or blacklisted stream can also be distributed through a multicast. In this scenario the bandwidth demand at the edge is constant regardless of the number of consumers, and is always less than or equal to the bandwidth required to stream the video with no privacy-preserving transformations.

However, a video stream will often have multiple consumers with different objectives, situated at different locations. Consider a smart-city scenario, where we have multiple consumer applications (or human viewers) utilizing the same camera feed. Table \ref{tab:amadeus-scenario} showcases five smart-city applications, each responsible for analyzing different types of objects in the video stream, along with the object classes that need to be disclosed or redacted based on whether whitelisting or blacklisting is used. In this scenario, a naive approach involves creating consumer-specific video streams to satisfy each consumer's requirements while also preserving privacy. This can cause the bandwidth requirement at the edge to rapidly increase as multiple consumers are added. 



\begin{figure}[t]
	\centering
	\includegraphics[width=\columnwidth]{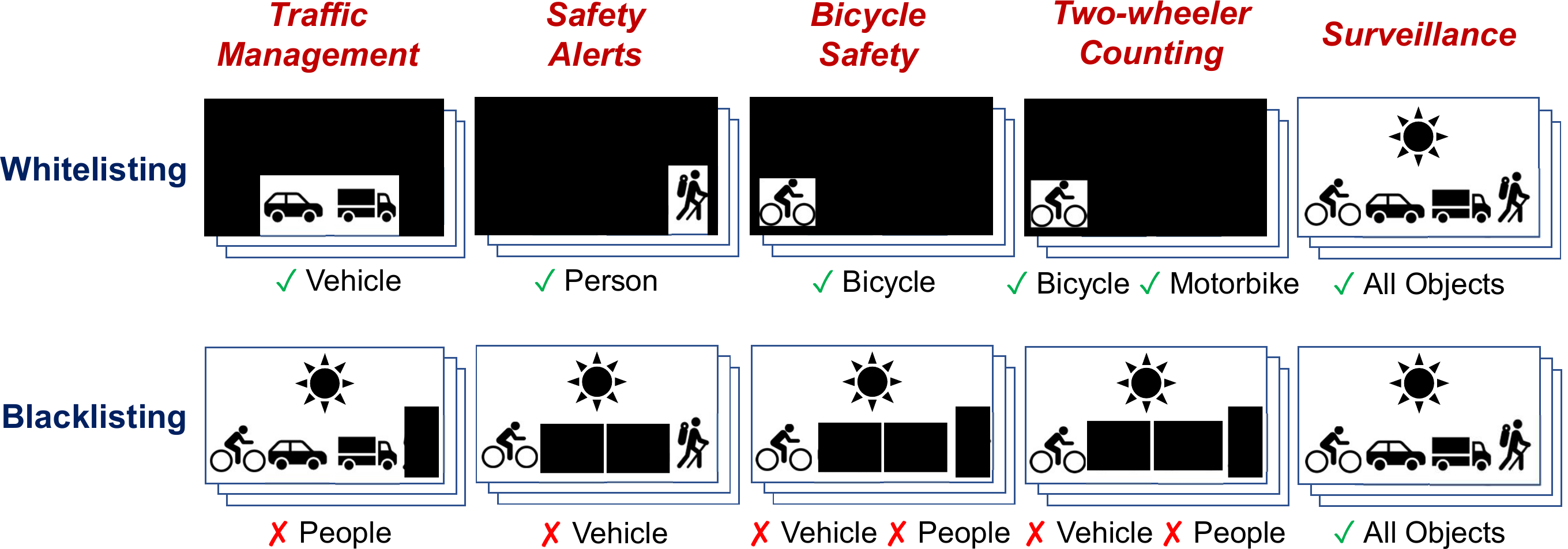} 
	\vspace{-0.6cm}
	\caption{Multi-consumer Smart-City Scenario}
	\label{fig:smart-city}
\end{figure}

Consider the use of blacklisting in the smart-city scenario. Figure \ref{fig:smart-city} illustrates the five \textit{blacklisted} video streams created to satisfy each application, Each of these streams contain a common background, which in this case is streamed five times from the edge. In the adverse setting, when no objects are detected as belonging to any of the blacklisted classes, each consumer application gets the same frame in its entirety. 

If we use whitelisting to preserve privacy, and create a unique stream for each application, then some object classes may be in two or more streams. For example, as shown in Figure \ref{fig:smart-city}, objects belonging to the class \textsf{bicycle} are used by both the bicycle safety and two-wheeler counting applications. Thus, these objects are transmitted twice from the edge. 
However, this is still more efficient than blacklisting, as the background is blacked out and not disclosed to the viewers. 

\textbf{Observation 4:} Video frames are typically encoded and transmitted in a compressed format such as Motion JPEG \cite{mjpeg} or H264 \cite{h264}. These encoding formats encode information in the frequency domain. Hence, blacked-out regions do not contain any information or contribute to streaming bandwidth. 

\textbf{Streaming Objectives:} Consider a pixel $\rho_{i,j}^t$ of a frame $t$ as the smallest unit of information in a video stream with resolution $(W,H)$, where $0<i<W, 0<j<H$ and $t,W,H \in Z^+$. By using an object detector, if a pixel lies inside the bounding box for an object, we can classify it as: (i) belonging to that object class, else (ii) as belonging to the background. Therefore, to distribute the video stream in a privacy-preserving and bandwidth-efficient manner we need to do the following:
\begin{itemize}[noitemsep]
	\item Each pixel $\rho_{i,j}^t$ in a frame $t$ of the video stream contains useful information. Therefore, to communicate all the useful information while minimizing bandwidth, each pixel in the frame must be encoded only once.
	\item Each pixel $\rho_{i,j}^t$ in the frame $t$ must only be accessible to consumers who are authorized to view it, i.e., if it belongs to a \textit{whitelisted} class $R\in \omega_k$ or does not belong to a \textit{blacklisted} class $S\in \beta_k$ for the consumer $C_k$.
\end{itemize}

An \textbf{Ideal Solution} to achieve both objectives is as follows:
\setlist[enumerate]{noitemsep, nolistsep}
\begin{enumerate}[noitemsep]
	\item Assign each pixel $\rho_{i,j}^t$ in frame $t$ to a set $\Gamma_c$, where $c \in \{\Omega, background\}$, and $\Omega$ represents all the object classes which an object detector can detect.
	\item Encrypt each pixel set $\Gamma_c$ with a unique key $\kappa_c$.
	\item Encode all the encrypted pixels $\rho_{i,j}^{t}{'}$ as a single frame.
\end{enumerate}

Using this \textit{ideal} scheme, we could in theory distribute this single encrypted video stream to multiple consumers. 
However, because each set of pixels $\Gamma_c$ is encrypted with its own key $\kappa_c$, \textit{only} consumers with the correct key will be able to decrypt and decode the corresponding pixels. 

\textbf{Key Properties:} This scheme effectively converts the video-stream distribution problem into one of key management. 
Combining this with the knowledge of a consumer's whitelisted classes, we can authenticate each consumer and \textit{only} provide them the set of keys they require to view their relevant sets of pixels, thus:
(i) \textbf{preserving privacy} as each viewer can only \textit{see} objects belonging to their whitelisted classes, 
(ii) \textbf{preserving bandwidth} as each pixel is only transmitted once from the edge, and 
(iii) \textbf{reducing computational costs} as each pixel is only encrypted and encoded once.
  
Although not a natural fit, we can also use the above-mentioned approach to generate a blacklisted stream, by providing a consumer $C_k$ all the keys $\kappa_c$, for all classes $\forall c\notin \beta_k$. 

\textbf{Video-Streaming Realities}: Unlike the proposed scheme, video-streaming techniques first perform encoding, and then encrypt the encoded data \cite{h264, hls, mpeg-dash}, as performing encryption on raw frames 
destroys \textit{redundant} information in a video frame, which is used to efficiently compress and encode the video. Therefore, we propose a practical scheme called \system which performs encoding followed by encryption.

\begin{figure}[t]
	\centering
	\includegraphics[width=\columnwidth]{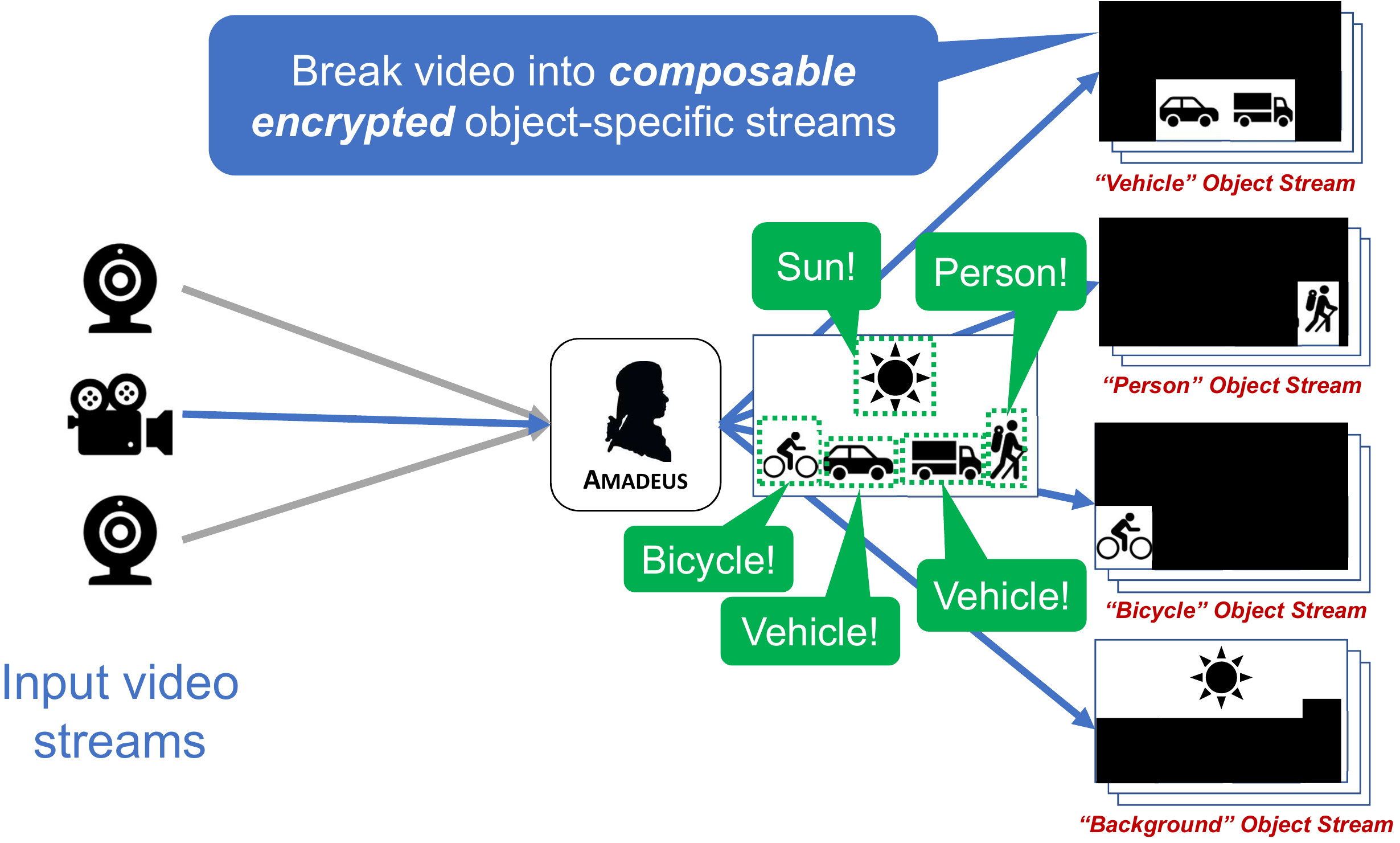} 
	\vspace{-0.6cm}
	\caption{\system: \textit{Composable} Encrypted Streaming}
	\label{fig:amadeus}
\end{figure}

\textbf{\system} creates $N+1$ encrypted video streams for each object class $c \in \{\Omega, background\}$, where $\Omega$ represents all the object classes which an object detector can detect, and $N$ is the number of object classes in $\Omega$. We call these \textit{object streams}. Each object stream contains all the pixels $\rho_{i,j}^t \in \Gamma_c$ corresponding to objects detected as belonging to a particular class $c$, and is encrypted with an object-class-specific key $\kappa_c$. $N$ can be dynamically changed based on the object classes we need to detect, based on application white/blacklisting requirements. \system also creates a residual background stream, which contains all the pixels not assigned to any object class. This background stream is encrypted with a special key $\kappa_{background}$. Figure \ref{fig:amadeus} provides an overview of \system considering the multi-application smart-city example. 

\textbf{Composable Streams}: Each pixel is assigned to \textit{only} one of the $N+1$ created object streams. As shown in Figure \ref{fig:amadeus}, all the areas of an object stream not assigned any pixels are blacked out and do not contain any information. Therefore, post decryption, we can combine all the $N+1$ streams, using simple pixel addition, to reconstruct the original video stream. 

\textbf{Compute + Bandwidth Efficiency:} \system still satisfies the key desirable properties provided by the \textit{ideal scheme}. For instance, each object stream is encrypted with a different key. Therefore, while every user can consume all the object streams through a Content-Distribution Network (CDN), they are \textit{only} able to decrypt the streams which they are authorized to view (whitelisting). Additionally, each pixel in a frame is assigned to only one object stream. Therefore, as each pixel is encoded, encrypted and transmitted to the CDN only once, \system can provide privacy-preserving video streaming in a compute and bandwidth-efficient manner. \system can also directly live-stream video to multiple consumers. However, in the absence of a CDN each consumer creates a direct streaming connection to the edge, which can cause bandwidth requirements at the edge to increase as new consumers are added. While, \system is bandwidth efficient at the edge, it can also save downstream bandwidth at the video consumer, as a consumer can now only consume relevant object streams instead of the entire video stream.

\textbf{Access Control:} \system can also generate new encryption keys at periodic intervals to ensure better security, while allowing consumer viewing permissions to change over time. An additional benefit of \system is the fact that despite having a subset of all the encryption keys $\kappa_c$, every consumer can still stream and locally store the encrypted video stream in its entirety. This opens up the possibility of a consumer obtaining permission to process a previously \textit{inaccessible} object stream from the administrator in the future. A good example of this can be surveillance video. We can envision a situation where security personnel can view all aspects of a video frame except human faces. However, if a crime is committed, they can request access to the encryption key $\kappa_{faces}$ required to view faces in a particular time duration, in order to analyze the archived footage. 
This process is comparable to a legal search warrant, where a judge needs to grant permission to perform an investigative search.

\subsection{Administrative Benefits of Whitelisting} \label{sec:administrative}
We now describe the benefits of whitelisting from the standpoint of administering privacy-preserving live streams to multiple consumer applications.

Blacklisting requires objects belonging to \textit{sensitive} classes to be redacted from the consumer's video stream. Therefore, the video-stream administrator has to anticipate: (i) which object classes are sensitive and can disclose privacy, and (ii) whether redacting sensitive objects impacts the performance of a consumer application. Therefore, in this \textit{administrator-centric} permissions model, an administrator needs to balance both user privacy and application utility.

Whitelisting requires objects belonging to \textit{relevant} classes to be disclosed to a consumer. Therefore, consumers are asked to specify what object classes they require to meet their stated objective. These requests can then be approved or declined by the video-stream administrator. This is similar to the permissions model used in smartphones \cite{smartphone-permissions}, where applications must request the smartphone user to grant access to a certain set of capabilities. Thus, whitelisting features a \textit{consumer-centric} permission model. This is significantly easier to administer as it shift the onus of \textit{specifying} utility requirements on the consumer, while the administrator checks if application requirements are invasive of privacy. 

However, \system authorizes applications at the granularity of the object classes supported by the object detector. Therefore, an administrator must \textit{presume} that if an object stream corresponding to an object class is disclosed to an application, then it can perform any operation on those objects.

\system can be implemented using a proprietary format. However, to ensure compatibility with real-world systems, our implementation uses existing streaming formats. This introduces some overheads. We describe this implementation and benchmark these overheads in subsequent sections. We believe that the key ideas of \system are also applicable to other content streaming domains, including text and audio.

\section{\system Implementation} \label{sec:implementation}
\begin{figure}[t]
	\centering
	\includegraphics[width=\columnwidth]{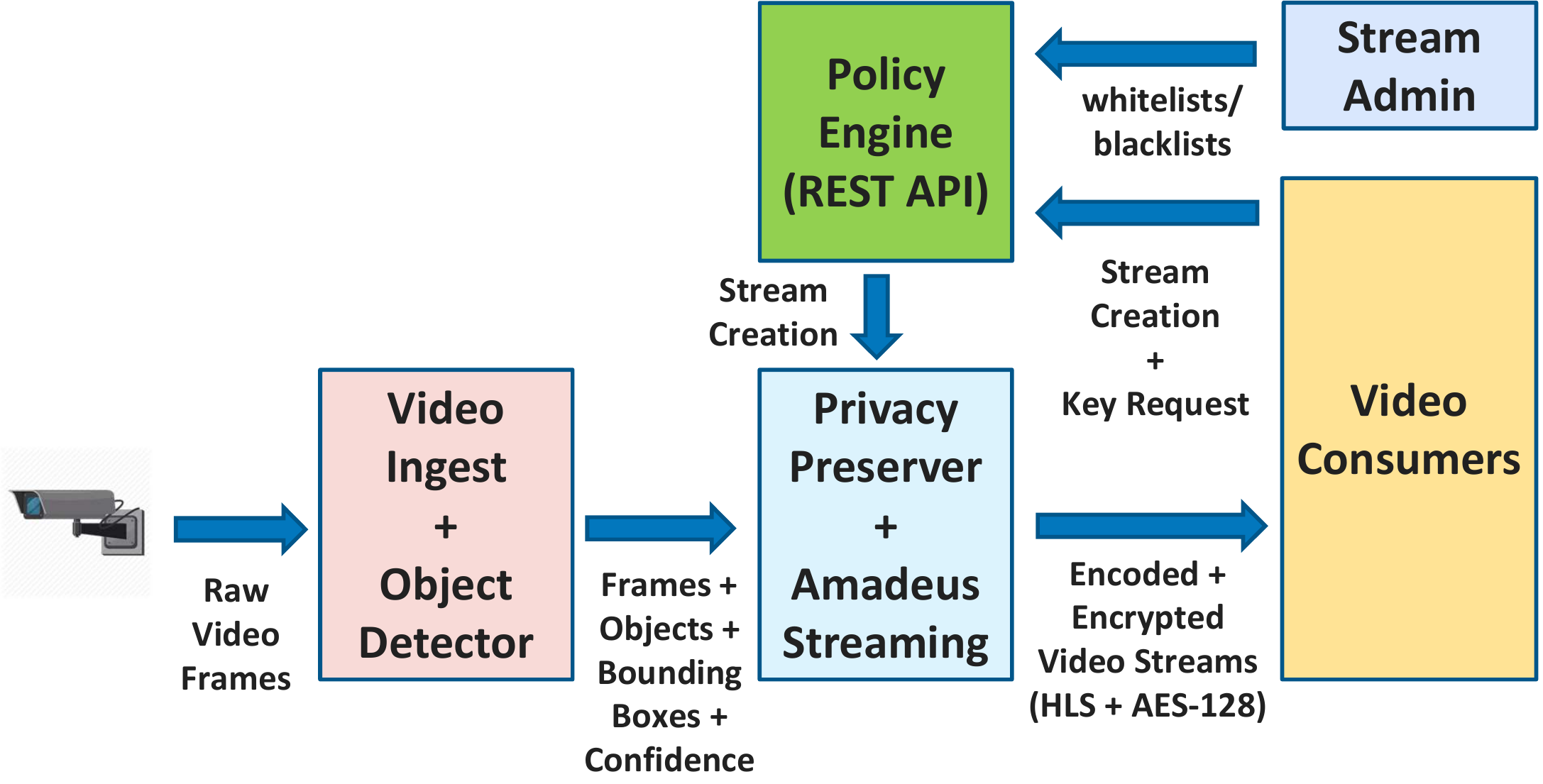} 
	\vspace{-0.6cm}
	\caption{\system: Architecture}
	\label{fig:defocus-arch}
\end{figure}

Our realization of the \system privacy-preserving live streaming and analytics pipeline, as visualized in Figure \ref{fig:defocus-arch}, builds on existing video-streaming technologies. This makes it compatible with existing applications and CDNs. We now describe each of \system' component modules. 
 
%
%
%

\subsection{Policy Engine} \label{sec:policy-engine}
The policy engine exposes both a consumer and administrator-facing REST API \cite{rest}. 

\textbf{Administrator API:} An administrator is allowed to create/remove consumer profiles, as well as specify object classes that a consumer is allowed to view (whitelisting), or not allowed to view (blacklisting). 

\textbf{Consumer API:} The consumer API allows consumers to request to view a set of object streams, by specifying a list of object classes. The request is denied if any of the requested classes are: (i) \textit{not} in the whitelist, 
or 
(ii) in the blacklist 

\textbf{Key Management:} The policy engine is also responsible for generating, revoking and distributing the encryption keys used to protect the object streams. 

\subsection{Object Detector} \label{sec:object-detector}
The object-detection module is responsible for ingesting a raw video feed, and detecting the objects in each frame. 
Currently, we utilize Keras-based \cite{keras} implementations of Faster-RCNN \cite{faster-rcnn}, YOLOv3 \cite{yolo-v3} and Tiny-YOLOv3 \cite{yolo-v3}. 

As described in Section \ref{sec:tracking}, the object-detection module also supports running detection only once every $n$ frames, followed by object tracking. We utilize the SORT \cite{sort} framework to create and keep track of multiple objects, in conjunction with object trackers \cite{opencv-trackers-1, opencv-trackers-2} available in OpenCV\cite{opencv}. 


\subsection{Privacy Preserver} \label{sec:privacy-preserver}
The privacy preserver is responsible for implementing the \system streaming mechanism, and generates composable encrypted object streams. These streams are created based on the object-detection data received from the object detector. 
 
Instead of generating streams for all the object classes which a detector can detect, the privacy-preserver relies on the policy engine to dynamically specify which object classes consumers (or administrators for blacklisting) are interested in. 
All pixels not detected as belonging to any class specified by the policy engine are encoded into the background stream. 

\textbf{Nested Objects:} When using bounding boxes to disclose/redact objects, there is a possibility of some objects being nested within or overlapping with other objects. For example, if the object detector being used can detect both \textsf{faces} and \textsf{persons}, the bounding boxes for the \textsf{faces} will most likely overlap or be nested within the bounding box corresponding to the \textsf{person}. As described in Section \ref{sec:amadeus}, each pixel is assigned to \textit{only} one of the object streams. Therefore, to prevent privacy loss, we assign overlapping pixels to the \textit{smaller} bounding box. In the described scenario, this ensures that \textsf{faces} get assigned to their own object stream, and consequently redacted from the \textsf{person} stream.

\textbf{Live Streaming:} The privacy preserver supports two types of outputs for each object stream: (i) raw frames to be locally consumed by applications on the edge, and (ii) H264-encoded \cite{h264} HTTP Live Streams (HLS) \cite{hls} 
for consumption by external consumers. HLS encodes a video as files of a fixed-configurable duration, and uses HTTP to transport them. This file-based nature of HLS makes it especially useful from a content-distribution standpoint. 
Each object stream is encrypted using AES-128 \cite{aes-128}, using the 128-bit per-object-stream encryption keys generated by the policy engine.  

\textbf{Metadata Streaming:} 
\system also streams encrypted metadata corresponding to every object stream. This per-frame metadata contains a list of detected objects corresponding to the object class, their bounding boxes and confidence scores. This is useful as: (i) metadata providing \textit{live analytics} may be sufficient for many applications, and (ii) low-confidence object detections may be withheld from the user to prevent privacy loss. An application can then use metadata to detect the presence of such objects. If required, an application can \textit{recover utility}, by requesting that the frame be re-analyzed by a specialized object detector. 




\section{Evaluation} \label{sec:evaluation}
We evaluate \system using real-world video streams and applications, and investigate the 
(i) 
\textbf{frame rate} at which \system can stream privacy-preserving video, 
(ii) \textbf{privacy} and \textbf{utility loss} that \textit{whitelisting} and \textit{blacklisting} yield, 
(iii) \textbf{bandwidth} required by \system to stream privacy-preserving video from the edge, and
(iv) \system's \textbf{usability} measured from the perspective of real-world applications.

Some of our key results are as follows:
\begin{enumerate}
	\item \system' whitelisting approach can yield up to \textasciitilde5000x lower privacy loss as compared to blacklisting.
	\item \system reduces the bandwidth required to stream video from the edge by up to \textasciitilde5.5x as compared to the naive blacklisting approach.
	\item \system is usable by real-world applications and yields up to \textasciitilde28x lower privacy loss.
\end{enumerate}


\begin{table}[t]
\centering
\small
\caption{Evaluation Video Data Description}
\vspace{-0.2cm}
\label{tab:videos}
\resizebox{\columnwidth}{!}{%
	\begin{threeparttable}
		\begin{tabular}{c|l|l}
			\toprule
			\textbf{Video} & \textbf{Camera} & \textbf{Description}  \\ 
			\midrule
			chicago-urban  & moving & Chicago urban scenes \\
			\midrule
			mumbai-urban   & moving & Mumbai urban scenes  \\
			\midrule
			new-york-urban & moving & New York City urban scenes   \\
			\midrule
			abc-action-news  & moving & news feed about traffic cameras  \\
			\midrule
			abc-7-news      & moving & news feed about traffic rules  \\
			\midrule
			cbs-la-news      & moving & news feed about traffic collisions \\
			\midrule
			jackson-hole-town & static & Jackson Hole town-square camera \\
			\midrule
			jackson-hole-restaurant & static & surveillance camera outside a restaurant \\
			\midrule
			m4-motorway & static & traffic camera on the M4 highway (UK)\\
			\midrule
			seattle-urban-1 & static & residential area in Seattle \\
			\midrule
			seattle-urban-2 & static & university intersection in Seattle \\
			\midrule
			seattle-urban-3 & static & university intersection in Seattle \\			
		\end{tabular}
	\end{threeparttable}}
\end{table}

We deploy \system on a mid-range machine with an 8-core Intel i7-7700 processor \cite{intel-i7-7700}, 16 GB of memory and an Nvidia GTX 1070 \cite{gtx-1070} GPU, which is comparable to the Nvidia Xavier edge platform \cite{nvidia-xavier}. 
We perform experiments using 12 videos described in Table \ref{tab:videos}, which include static traffic cameras, urban scenes captured by moving cameras, and news-reel footage. The videos in our dataset each contain multiple object classes (including vehicles, people and bicycles), and were recorded at 25 or 30 FPS. 


\textbf{Applications:} We utilize two real-world applications: (i) a background-subtraction-based vehicle counting application \cite{car-counting} and (ii) a face-detection application which also predicts gender and age for each detected face \cite{face-detection}. While we utilize both the vehicle-counting and face-detection application to measure utility loss (the number vehicles/faces which were not counted), we also utilize the face-detection application in an adversarial setting to measure privacy loss (the number of faces that were detected when faces were not shown or redacted from the consumer).

\subsection{End-to-End System Measurements}
\system is 
deployed at the edge, and hence it needs to deliver a relatively-high video frame rate while meeting edge resource constraints. We now benchmark the resources used by \system and the end-to-end frame rate it can deliver. 

\textit{End-to-end} performance involves the entire pipeline including: object detection, privacy-preserving video whitelisting/blacklisting, and video encoding and encryption. Therefore, it is a function of the following variables: 

1) \textbf{Video Resolution}, which determines the number of pixels to process. Therefore, we consider two video streams: \textsf{jackson-hole-town} with a full-HD (1920x1080) resolution, and \textsf{abc-action-news} with an HD (1280x720) resolution.
	
2) \textbf{Object-detection} complexity which influences system requirements. Therefore, we consider (i) a state-of-the-art object detector: YOLOv3 (\textsf{yolo}) \cite{yolo-v3}, (ii) a low-cost object detector: Tiny-YOLOv3 (\textsf{tiny-yolo}) \cite{yolo-v3}, and (iii) performing object detection using YOLOv3  every $n=10$ frames (\textsf{yolo-skip-10}) followed by object tracking using the \textsf{SORT} \cite{sort} framework and an OpenCV-based KCF tracker \cite{opencv-trackers-1}. 

3) \textbf{System configuration} limits the frame rate which can be achieved. Therefore, we compare performance between running object detection on the GPU and the CPU.

4) \textbf{Video encoding and encryption} complexity, which depends on the number of output object streams that have to be created. Therefore, we vary the number of object streams required to service multiple consumer applications. 
	
\begin{figure}[t]
	\centering
	\includegraphics[width=\columnwidth]{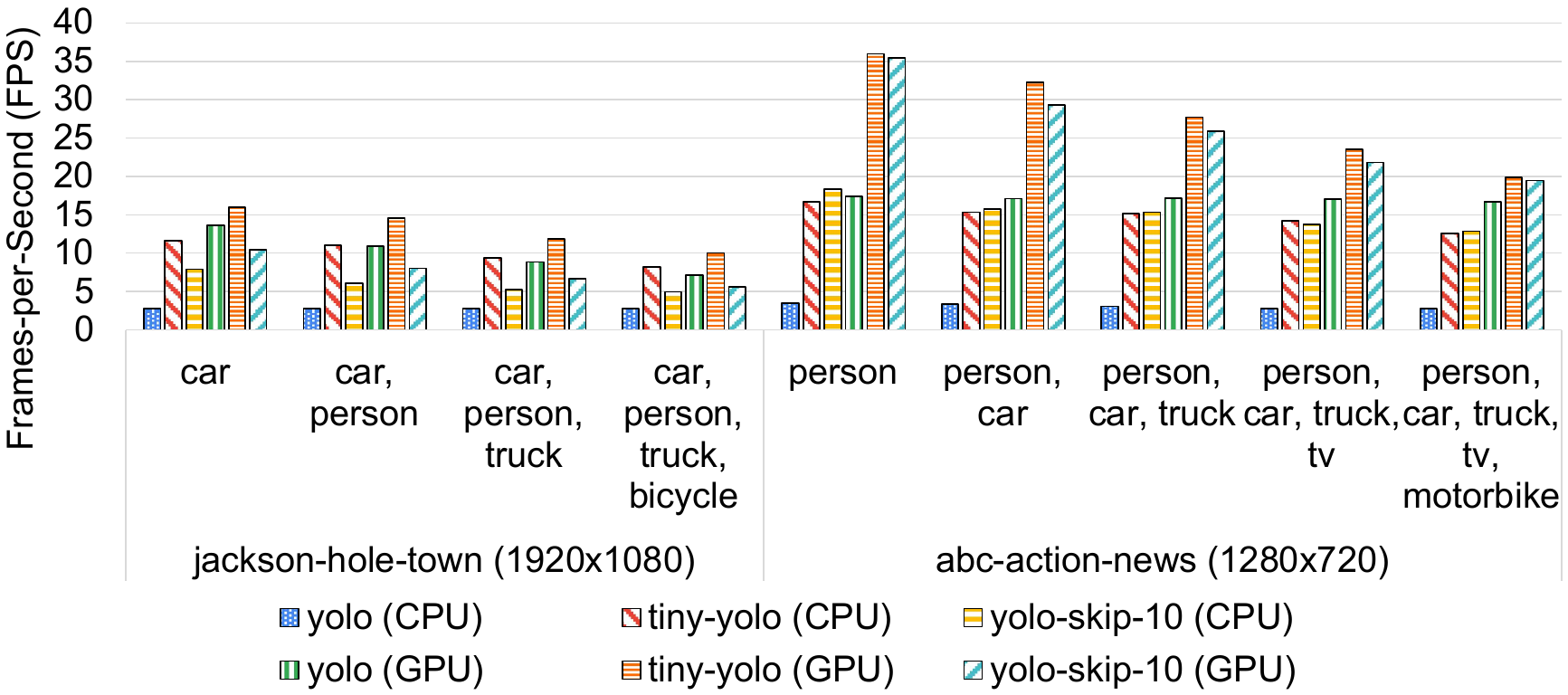} 
	\vspace{-0.6cm}
	\caption{\system measured end-to-end frame-rate} 
	\label{fig:end2end-fps}
\end{figure}

\begin{figure}[t]
	\centering
	\includegraphics[width=\columnwidth]{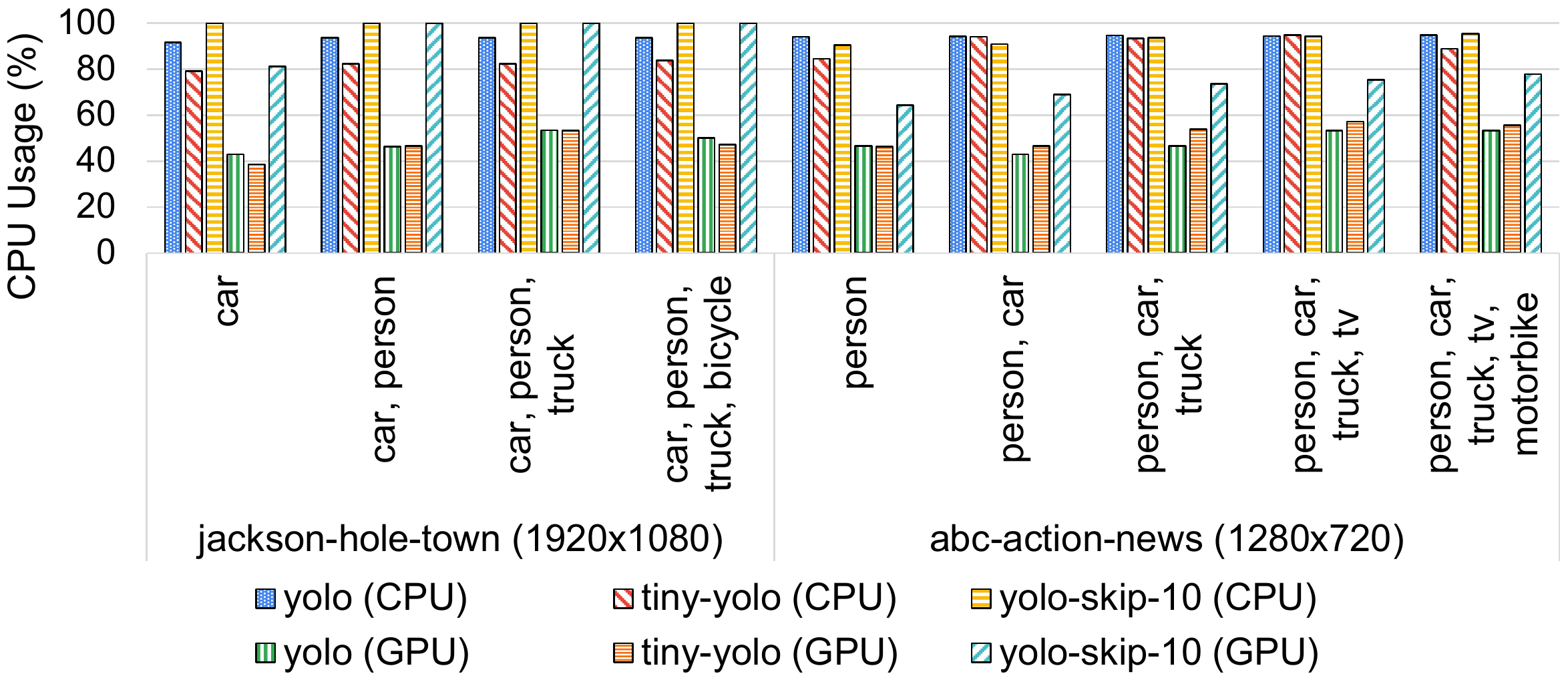} 
	\vspace{-0.6cm}
	\caption{\system measured CPU utilization} 
	\label{fig:cpu-usage}
\end{figure}

Figures \ref{fig:end2end-fps} and \ref{fig:cpu-usage} plot \system's output frame rate, and CPU usage respectively, for two videos, \textsf{jackson-hole-town} and \textsf{abc-action-news}, while varying the (i) object detectors, (ii) object-detection deployment device (GPU or CPU), and (iii) the number of output object streams. In terms of object-stream ordering we start with the highest-occurring object class in the video, and subsequently add streams in order of their occurrence. Note that in each case, all the pixels not belonging to a requested stream are also encoded into the \textit{background} stream. Therefore, the number of pixels encoded and encrypted in each case is constant. 
The key insights garnered from Figures \ref{fig:end2end-fps} and \ref{fig:cpu-usage} are:

\textbf{Platform:} As expected, running object detection on the GPU yields significantly higher frame rates than the CPU -- up to \textasciitilde5x for \textsf{yolo} and \textasciitilde2x for \textsf{yolo-skip-10} and \textsf{yolo-tiny}. 

\textbf{Object Detection:} We observe that using the low-cost \textsf{tiny-yolo} yields the highest end-to-end frame rate -- up to 2x higher than both \textsf{yolo} and \textsf{yolo-skip-10} on the GPU. Additionally, for \textsf{tiny-yolo}, the frame rate falls with an increase in object streams. Thus, indicating that video encoding is the bottleneck in this case. On the other hand, for large models like \textsf{yolo} running on the GPU, the end-to-end frame rate can remain constant as we add additional object streams
. In this case, the object-detection model is the bottleneck.

\textbf{Object Tracking:} While object detection can run on the GPU, the object tracker we use only runs on the CPU. Figure \ref{fig:tracking-fps} in Section \ref{sec:design}, indicated that \textsf{yolo-skip-10} (detection + tracking) can yield similar or higher frame rates than \textsf{tiny-yolo}. If we consider the added CPU workload of encoding and encryption, this observation still holds for the HD video \textsf{abc-action-news}. On the contrary, for the full-HD \textsf{jackson-hole-town}, \textsf{yolo-skip-10} surprisingly yields even lower frame rates than \textsf{yolo} -- up to \textasciitilde0.33x lower. 
We believe that this is due to: (i) a higher number of objects in the \textsf{jackson-hole-town} video leading to a large number of trackers, and (ii) the higher video resolution leading to higher CPU usage for encoding and encryption, which leaves fewer CPU resources for object tracking. 

\textbf{Impact of Object Streams:} Adding object streams caused the frame rate to fall, CPU usage to rise (Figure \ref{fig:cpu-usage}), and memory requirements to increase linearly. In the worst case, the total memory used never exceeded \textasciitilde45$\%$.

If we consider using \textsf{yolo-skip-10} for object detection, \system can stream privacy-preserving HD video at a usable rate of \textasciitilde14-16 FPS using the CPU. Thus, \system is usable on a platform without a GPU. Using the mid-tier GPU we were able to stream HD video at \textasciitilde20-35 FPS.

\subsection{Privacy and Utility Loss}

Preserving privacy while allowing applications to extract utility is \system' key objective. As described in Section \ref{sec:privacy-loss-guarantees}, object detection errors namely, false positives and false negatives, play a key role in influencing the amount of privacy and utility loss. 
Therefore, we now measure the privacy and utility loss yielded by using whitelisting and blacklisting, while utilizing different object detectors. 

We compare Tiny-YOLOv3 (\textsf{tiny-yolo}) \cite{yolo-v3}, and object detection every $n$ frames using Faster-RCNN (\textsf{rcnn-skip-n}) or YOLOv3 (\textsf{yolo-skip-n}) followed by object tracking for $n=5$ and $n=10$. We perform measurements over 10 unlabeled videos described in Table \ref{tab:videos}, 
and utilize labels generated by YOLOv3 \cite{yolo-v3} as the ground truth for \textsf{tiny-yolo} and \textsf{yolo-skip-n}, and Faster-RCNN \cite{faster-rcnn} as the ground truth for \textsf{rcnn-skip-n}.

\begin{figure}[t]
	\centering
	\includegraphics[width=\columnwidth]{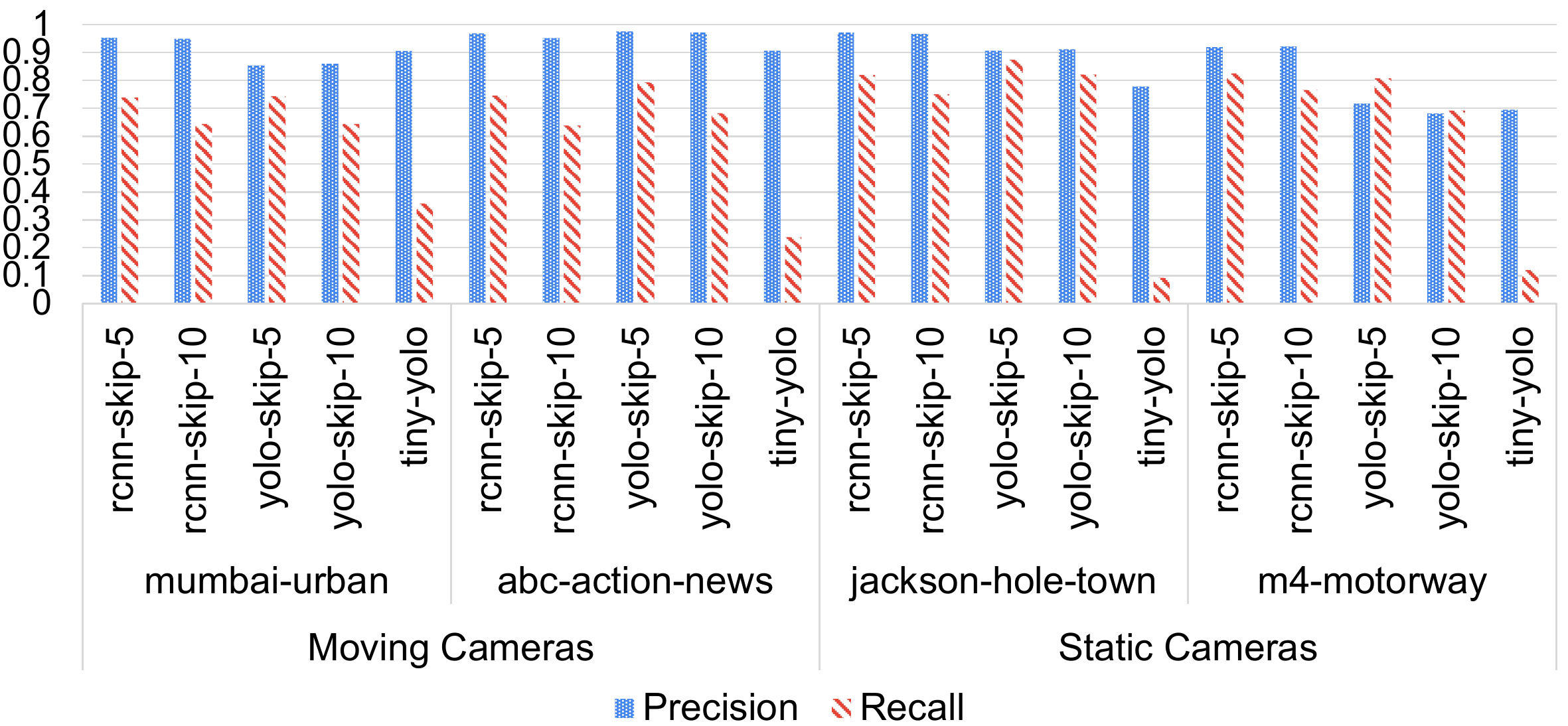} 
	\vspace{-0.5cm}
	\caption{Relative precision and recall measurements using different object-detection techniques}
	\label{fig:precision-recall-all-videos}
\end{figure}

\textbf{Precision and Recall:} As described in Section \ref{sec:implications-privacy}, precision and recall measure the prevalence of false positives and false negatives, i.e., higher precision translates to fewer false positives, and higher recall translates to fewer false negatives. For our test videos and object detectors, we observe that precision is high for all combinations, with an average precision of 0.91. If we consider individual videos, the precision is similar for techniques based on the same class of object detector. However, we observe an average recall of 0.65, but that recall steadily decreases as object-detection becomes \textit{lower cost}. Consider object-detection performed only once every $n$ frames followed by tracking. For both \textsf{rcnn-skip-n} and \textsf{yolo-skip-n} we observe that recall decreases as we go from $n=10$ to $n=5$. We also observe that \textsf{tiny-yolo} has the lowest recall (as low as 0.03). Figure \ref{fig:precision-recall-all-videos} plots the overall precision and recall of object detectors for 4 of the 10 videos.

\begin{figure}[t]
	\centering
	\includegraphics[width=\columnwidth]{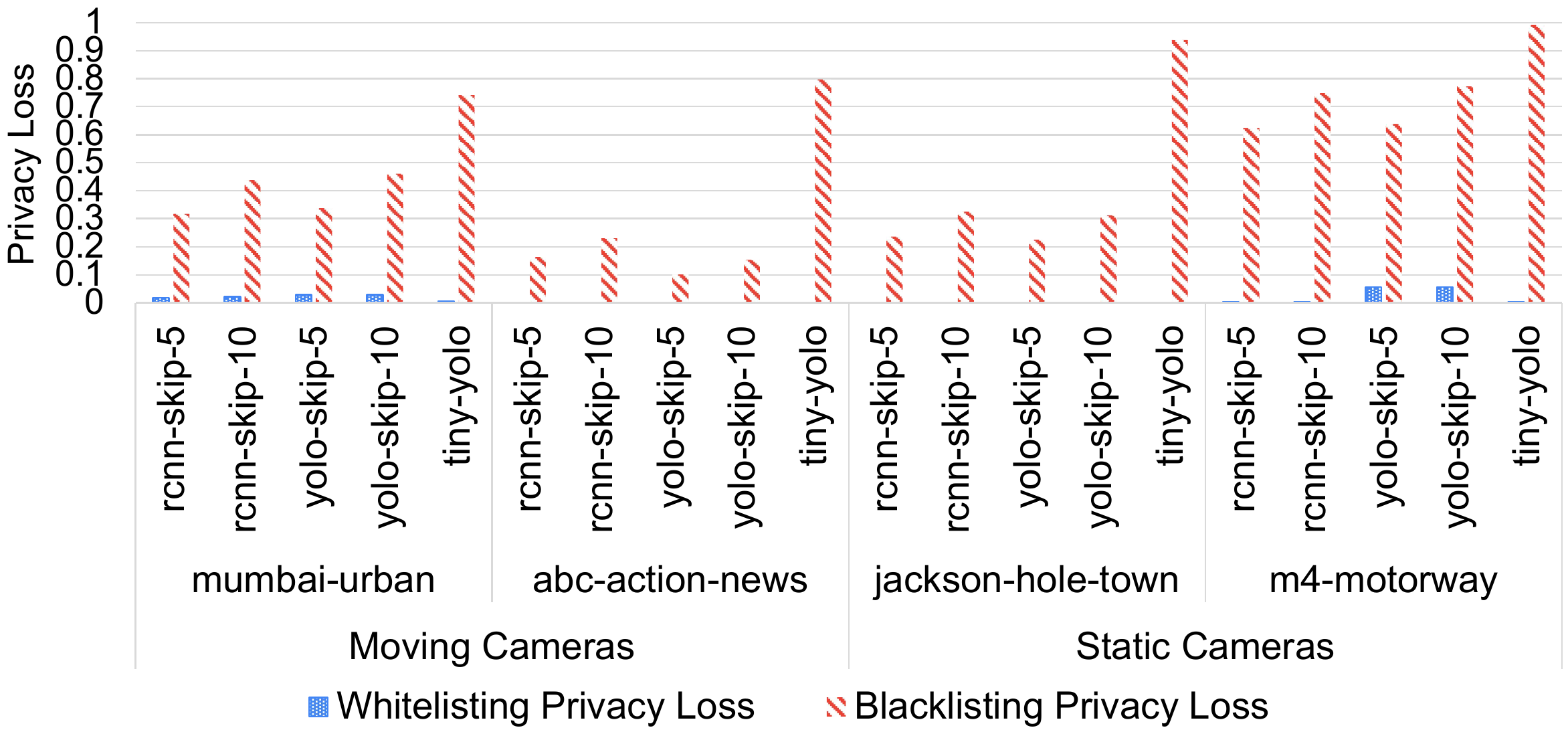} 
	\vspace{-0.5cm}
	\caption{Privacy-Loss Measurements}
	\label{fig:privacy-loss-all-videos}
\end{figure}

\begin{figure}[t]
	\centering
	\includegraphics[width=\columnwidth]{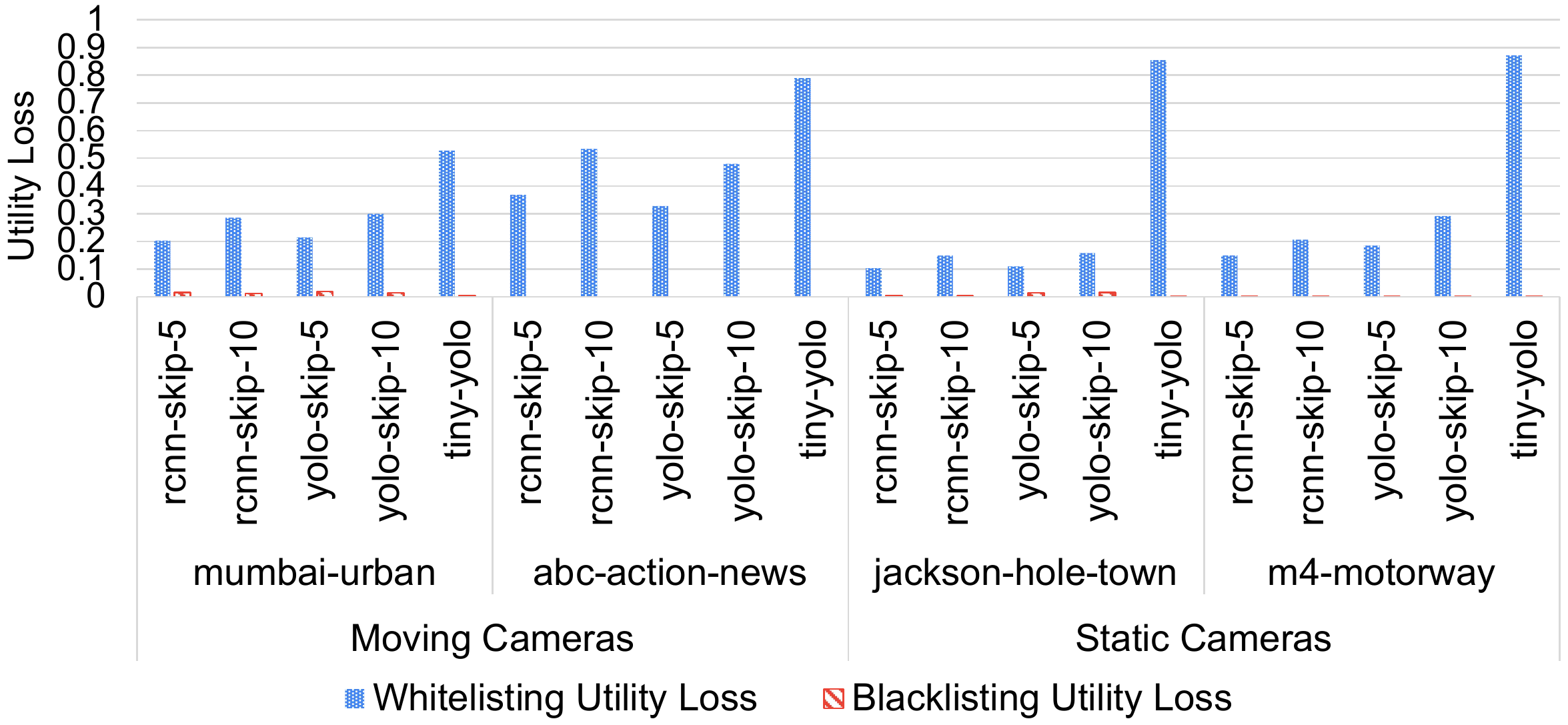} 
	\vspace{-0.6cm}
	\caption{Utility-Loss Measurements} 
	\label{fig:utility-loss-all-videos}
\end{figure}

\textbf{Privacy Loss:} Privacy loss depends on what classes are defined as sensitive. Consider a car-counting application. To successfully count cars, an application must be able to view all objects belonging to the class \textsf{car}. Therefore, whitelisting discloses only objects belonging to class \textsf{car}. On the other hand, blacklisting redacts objects belonging to sensitive classes. In this case, let the sensitive class be \textsf{person}. Therefore, while using whitelisting, all false positives classifying a \textsf{person} as a \textsf{car}, would cause privacy loss. On the other hand, for blacklisting, all false negatives (mis-detections) corresponding to the class \textsf{person}, would lead to a person not being redacted from the video stream. Figure \ref{fig:privacy-loss-all-videos} plots the potential privacy loss yielded by different object-detection techniques for 4 of the 10 videos we consider. Observe that whitelisting yields significantly lower privacy loss as compared to blacklisting -- up to \textasciitilde5000x fewer private objects disclosed in the worst case, and \textasciitilde72x fewer in the average case . 


\textbf{Utility Loss:} Figure \ref{fig:utility-loss-all-videos} plots the utility loss for different object detectors for 4 of the 10 streams. Whitelisting on average yields \textasciitilde34$\%$ utility loss compared to blacklisting's \textasciitilde0.6$\%$. 

We conclude that, while whitelisting on average yields \textasciitilde72x lower privacy loss, blacklisting yields \textasciitilde57x lower utility loss. However, while privacy loss is permanent, utility loss may be recoverable by a post-processing query. 
If we compare the low-cost \textsf{tiny-yolo} against \textsf{yolo-skip-10} (detection every 10 frames followed by tracking), we observe that \textsf{tiny-yolo} has both significantly higher privacy loss for blacklisting (up to \textasciitilde5x) and utility loss for whitelisting (up to \textasciitilde8x). Therefore, even though Figure \ref{fig:end2end-fps} indicates that both \textsf{tiny-yolo} and \textsf{yolo-skip-10} can deliver similar frame rates, \textsf{yolo-skip-10} is significantly better at preserving both privacy and utility.

\subsection{Bandwidth Measurements}
\system distributes privacy-preserving video to multiple consumers in a bandwidth-efficient manner. Therefore, we benchmark \system compared to the naive whitelisting and blacklisting approaches. For both \textit{naive} approaches, a custom video stream is created for each application. We perform experiments measuring bandwidth as a function of: (i) the number of consumer applications, and (ii) the number of object streams. The first experiment compares the bandwidth requirement of \system versus the naive approaches. As object detection plays a key role in determining which object stream each pixel in a frame is assigned to, the second experiment measures the impact of object-detection techniques on bandwidth, as we increase the number of object streams.

Bandwidth usage is also a function of frame rate, which depends on system configuration and object-detection techniques. Therefore, simply measuring the bandwidth used at the edge would not yield a fair comparison. Instead, we measure the total file size of the HLS-encoded video divided by the length of the video, and use it as a proxy for bandwidth. 

\begin{figure}[t]
	\centering
	\includegraphics[width=\columnwidth]{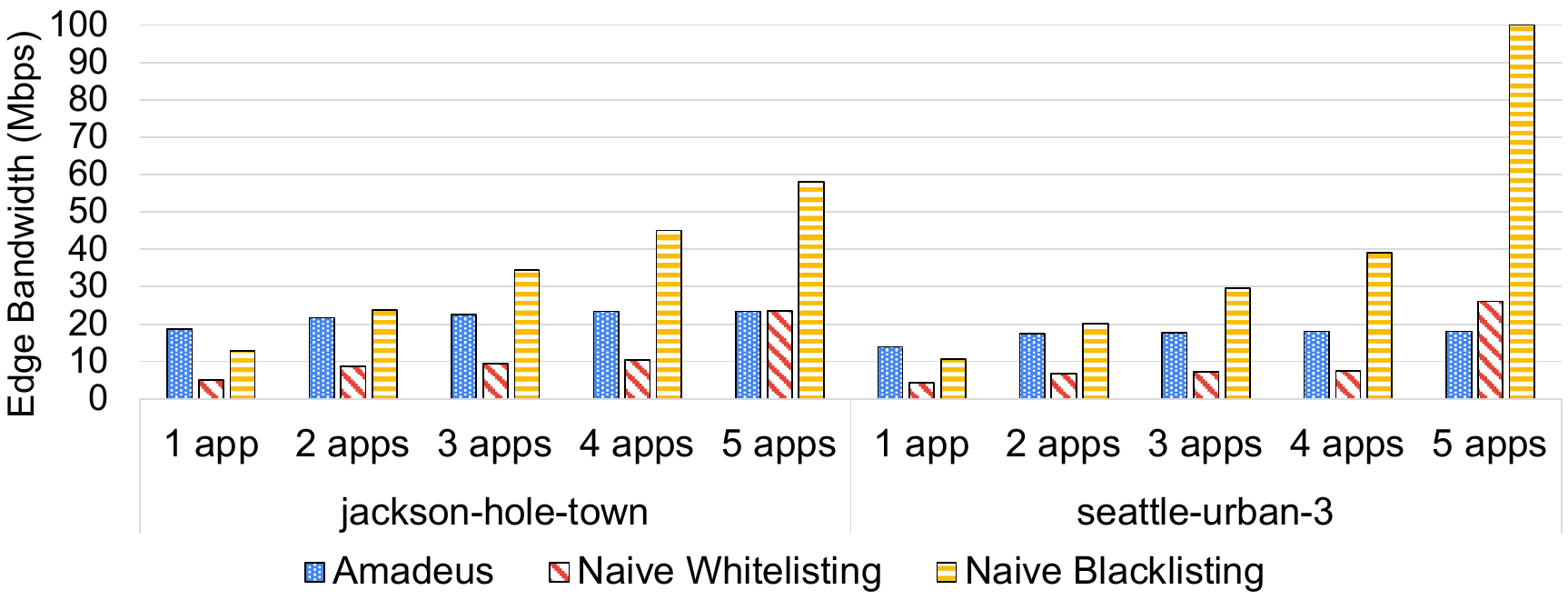}
	\vspace{-0.2cm}
	\caption{\system bandwidth requirements at the edge for serving multiple apps}
	\label{fig:bandwidth-app-scenario}
\end{figure}

\textbf{Multi-application Scenario:} We consider the five application smart-city scenario described in Table \ref{tab:amadeus-scenario} in Section \ref{sec:amadeus}.  
Figure \ref{fig:bandwidth-app-scenario} plots the output bandwidth (in Mbps) at the edge, for two different video streams, as we increase the number of consumer applications based on their ordering in Table \ref{tab:amadeus-scenario}.  
Observe that for a single application, both the naive approaches yield lower bandwidth -- up to 3.68x and 1.46x lower for whitelisting and blacklisting respectively. In this scenario both whitelisting and blacklisting are publishing a single stream with information \textit{removed}, as compared to \system which is \textit{re-encoding} all the data in the original stream, as separate object streams. However, as the number of applications increase, the naive blacklisting approach suffers significant increases in bandwidth requirement (\textasciitilde1.1-5.5x higher). This is because the naive blacklisting approach creates a separate stream per consumer, and each of these streams have the background pixels duplicated across them. On the other hand, as the number of applications increase, as compared to \system, the naive whitelisting approach yields lower bandwidth. This is because the whitelisted streams do not carry any background pixels. However, when we add the last application \textsf{Surveillance}, which requires an un-redacted stream, the naive whitelisting approach now needs to create a separate stream with the entire video. This causes the bandwidth requirements to rise up to 1.44x higher than \system. 

The key advantage of \system is that each pixel is encoded, encrypted and transmitted only once from the edge. However, using existing streaming technologies can increase overhead and \system can require up to 1.79x the bandwidth required for encoding the raw video stream. Additionally, as the number of object streams increase to support more applications, we see that \system' bandwidth requirements increase slightly. We believe that this is due to the overhead of per-stream headers and the breaking of the video stream into chunks by HLS. However, \system is still more efficient than the naive approaches, and the overhead can be reduced by creating a proprietary streaming standard.

\begin{figure}[t]
	\centering
	\includegraphics[width=\columnwidth]{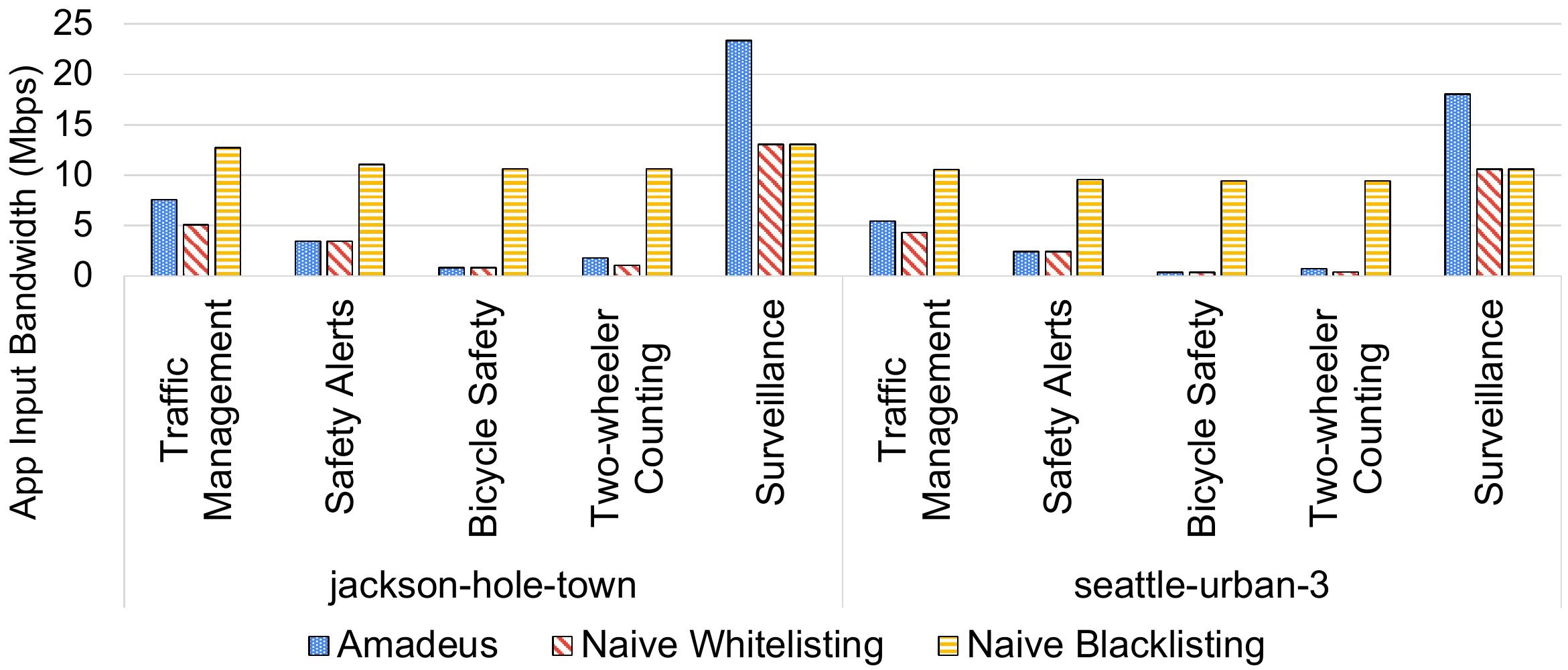}
	\vspace{-0.6cm}
	\caption{Input bandwidth at each app}
	\label{fig:bandwidth-app-input}
\end{figure}

We also measure input bandwidth at each application and the results are plotted in Figure \ref{fig:bandwidth-app-input}. Observe that \system can reduce input bandwidth at the application, as applications need not consume object streams not relevant to them. 

\begin{figure}[t]
	\centering
	\includegraphics[width=\columnwidth]{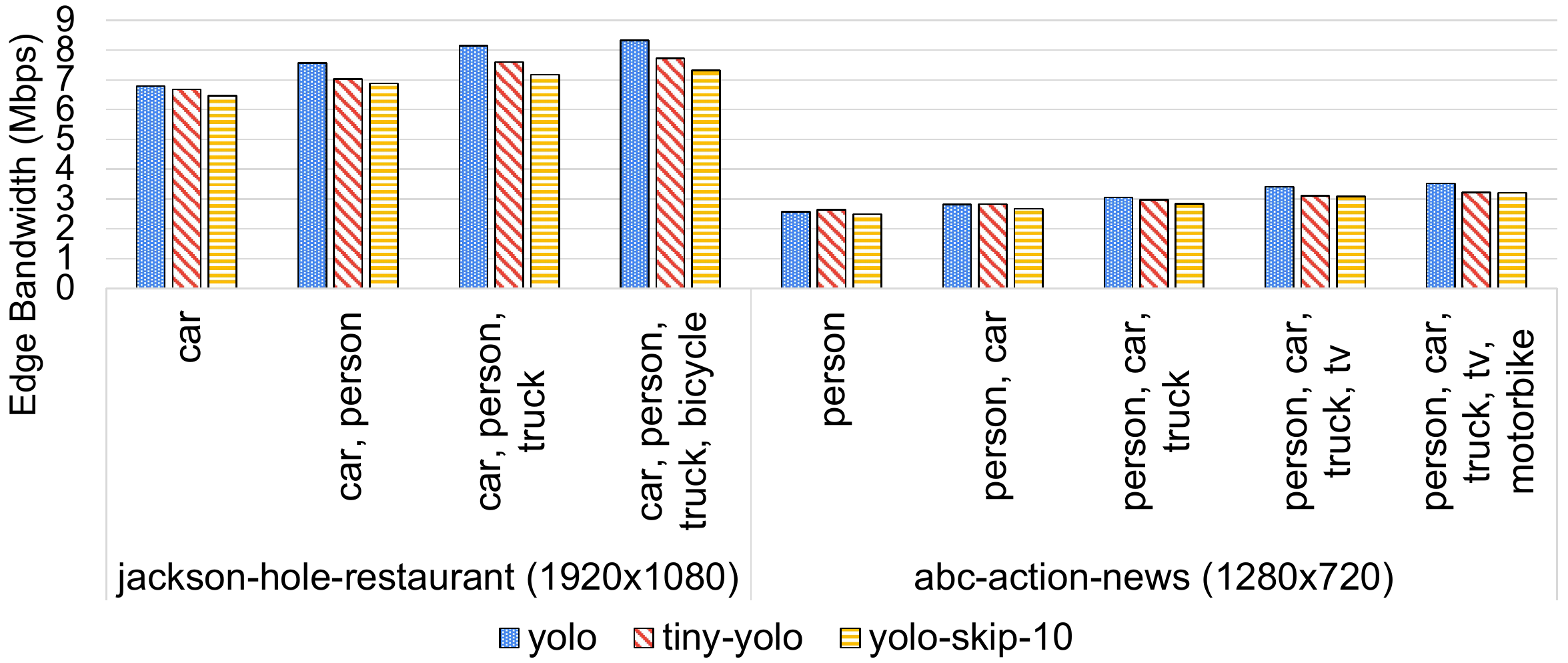} 
	\vspace{-0.5cm}
	\caption{\system bandwidth requirements at the edge for different object detectors}
	\label{fig:bandwidth-object-detection}
\end{figure}

\textbf{Impact of Object Detection:} Figure \ref{fig:bandwidth-object-detection} plots the bandwidth requirements for \system as the number of object streams increase, for different object-detection techniques: \textsf{yolo}, \textsf{tiny-yolo}, and \textsf{yolo-skip-10}. Note that each scenario also has a residual background stream. Due to the overhead of HLS, bandwidth requirements increase as we add new object streams. 
Observe that while using \textsf{tiny-yolo} or \textsf{yolo-skip-10}, the bandwidth requirements can be about \textasciitilde10-15$\%$ lower than \textsf{yolo}. This is because \textsf{yolo} can detect more objects, due to lower false negatives (Figure \ref{fig:precision-recall-all-videos}). This leads to more objects being moved to the object streams from the background.




We can conclude that while distributing video to multiple consumers, \system is bandwidth efficient at the edge, as when we add new applications there is minimal increase in bandwidth requirements. For instance, when adding the fifth application \textsf{Surveillance}, no additional bandwidth is required at the edge for \system. On the other hand the naive whitelisting and blacklisting approaches see \textasciitilde2.5x and \textasciitilde3.5x jump in bandwidth requirements respectively. This disparity between \system and the naive approaches can only get worse as more applications or object streams are added. 

\subsection{\system Usability Measurements}
The \textit{potential} privacy and utility loss calculated by equation \ref{eqn:privacy-utility-eqn} described in Section \ref{sec:privacy-loss-guarantees} provide a good idea of the impact of using blacklisted or whitelisted streams. However, while all objects may be perceptible by human viewers or custom detectors, they may not be perceivable by an application. Additionally, there is also a possibility of usability degradation (utility loss) due to (i) modification of the \textit{visual nature} of a video stream due to whitelisting/blacklisting, and/or (ii) model mismatch. Therefore, we measure privacy and utility loss by considering two real-world applications which utilize the \textit{composable} video streams generated by \system:
	
1) \textbf{Vehicle Counting:} This is an open-source background-subtraction-based vehicle-counting application \cite{car-counting}. It assumes a static camera, and uses background subtraction \cite{mog-background-subtractor} to detect moving objects. The application also takes as input, regions of the video frame, and counts all objects moving into those regions as \textsf{vehicles}. 
For this application to operate as intended, we whitelist all objects belonging to the classes \textsf{car}, \textsf{truck} and \textsf{bus}. In the blacklisting mode, we blacklist all objects belonging to the class \textsf{person}.
	
2) \textbf{Face Detection:} This is an open-source face-detection application \cite{face-detection} which also predicts gender and age for each detected face \cite{face-detection-model}.
For this application to operate as intended, we whitelist all objects belonging to the classes \textsf{person}.

Both the vehicle-counting and face-detection applications are used to measure utility loss, i.e., the number of vehicles/faces not detected. We also utilize the face-detection application as an \textit{adversary} to measure privacy loss, i.e., the number of faces that were detected when faces were not supposed to be disclosed. Since the videos we use are unlabeled, the detections on the un-redacted video is used as the baseline.

\textbf{Usability Experiments:} We compare three different scenarios against the baseline to measure utility loss for the vehicle-counting application where it consumes: (i) whitelisted video, (ii) a blacklisted video, and (iii) only the \textit{metadata}. Note that, as described in Section \ref{sec:privacy-preserver}, \system also provides a per-object-class metadata stream. This metadata stream contains the bounding boxes for all objects belonging to the class along with the detection confidence. 
By using the metadata stream corresponding to object classes \textsf{car}, \textsf{truck} and \textsf{bus}, the vehicle-counting application can directly count vehicles without the need to detect them using background subtraction. 
However, the face-detection app also tries to extract application-specific attributes like gender and age, which may not always exist in the metadata
, and hence, we measure utility loss only for the whitelisting case.

\begin{figure}[t]
	\centering
	\includegraphics[width=\columnwidth]{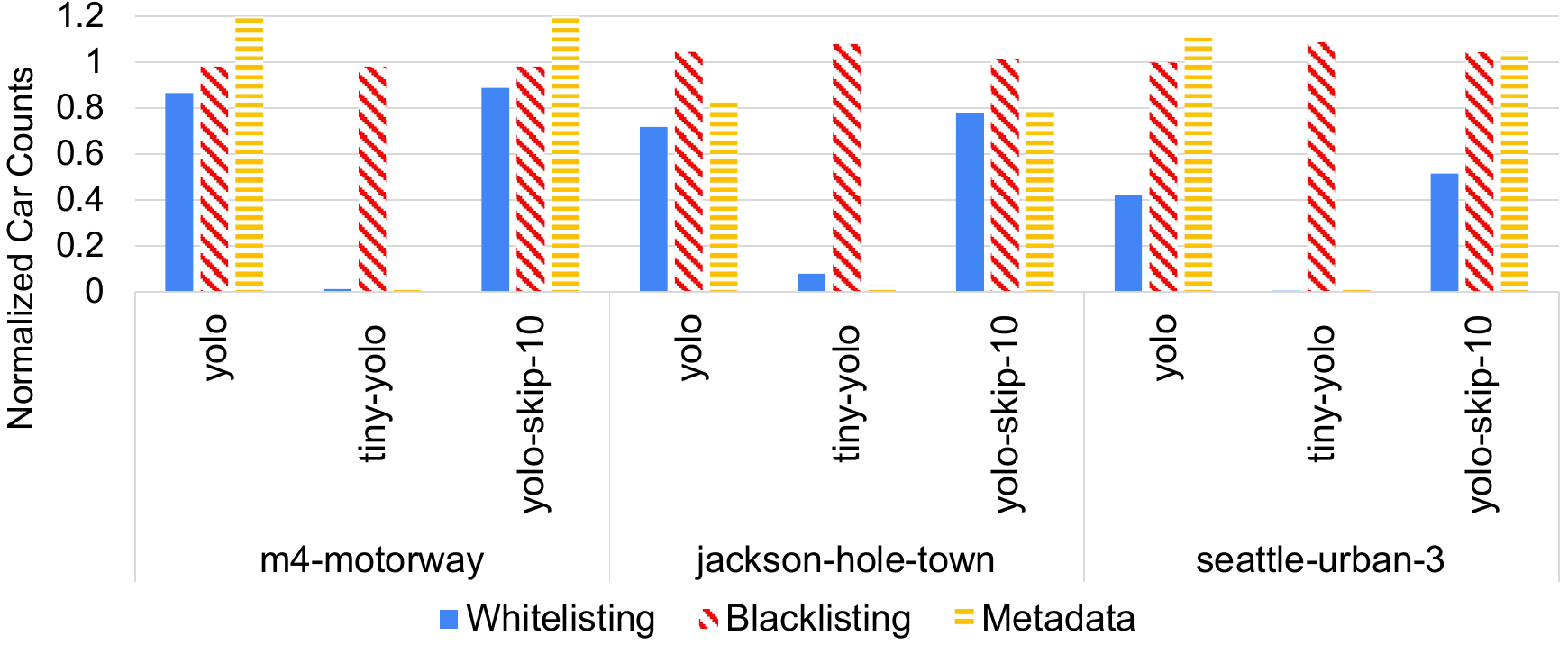} 
	\vspace{-0.6cm}
	\caption{Number of vehicles counted (normalized) by the vehicle-counting app.} 
	\label{fig:vehicle-detection-utility}
\end{figure}

Figure \ref{fig:vehicle-detection-utility} plots the number of cars that were counted by the application by utilizing the video/metadata streams generated by different object-detection techniques. We observe that the loss in utility is insignificant for blacklisting -- up to 2$\%$. On the other hand, for whitelisting, the utility loss is heavily dependent on the object-detection technique. For instance, while performing whitelisting using \textsf{tiny-yolo}, the utility loss is often more than 99$\%$. On the other hand, when using \textsf{yolo} and \textsf{yolo-skip-10} to generate the metadata, we can end up with higher car counts than the unmodified video stream. This is because the metadata already provides the detected objects of relevant classes (in this case vehicles), and the application need not perform object detection again. Thus, yielding better results than the background subtraction used by the application.

\begin{figure}[t]
	\centering
	\includegraphics[width=\columnwidth]{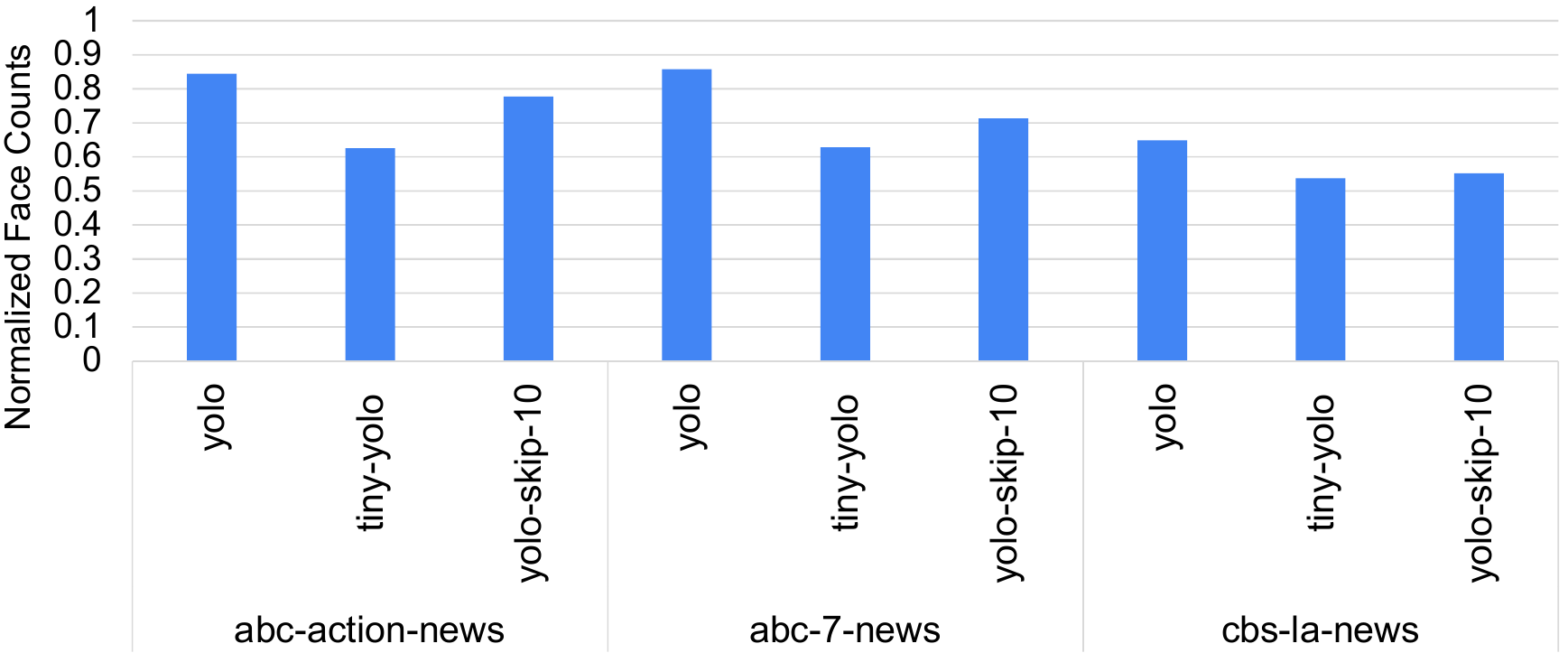} 
	\vspace{-0.3cm}
	\caption{Number of faces detected (normalized) by the face-detection app using whitelisting}
	\label{fig:face-detection-utility}
\end{figure}

We also measured the fraction of faces counted by the face-detection app using a whitelisted stream of objects detected as belonging to the class \textsf{person}. We observe that while using (i) \textsf{yolo}, the app detects \textasciitilde80$\%$ of faces, (ii) \textsf{tiny-yolo} the app detects \textasciitilde60$\%$ of faces, and (iii) \textsf{yolo-skip-10} the app detects \textasciitilde70$\%$ of faces. Figure \ref{fig:face-detection-utility} plots the results.

\begin{figure}[t]
	\centering
	\includegraphics[width=\columnwidth]{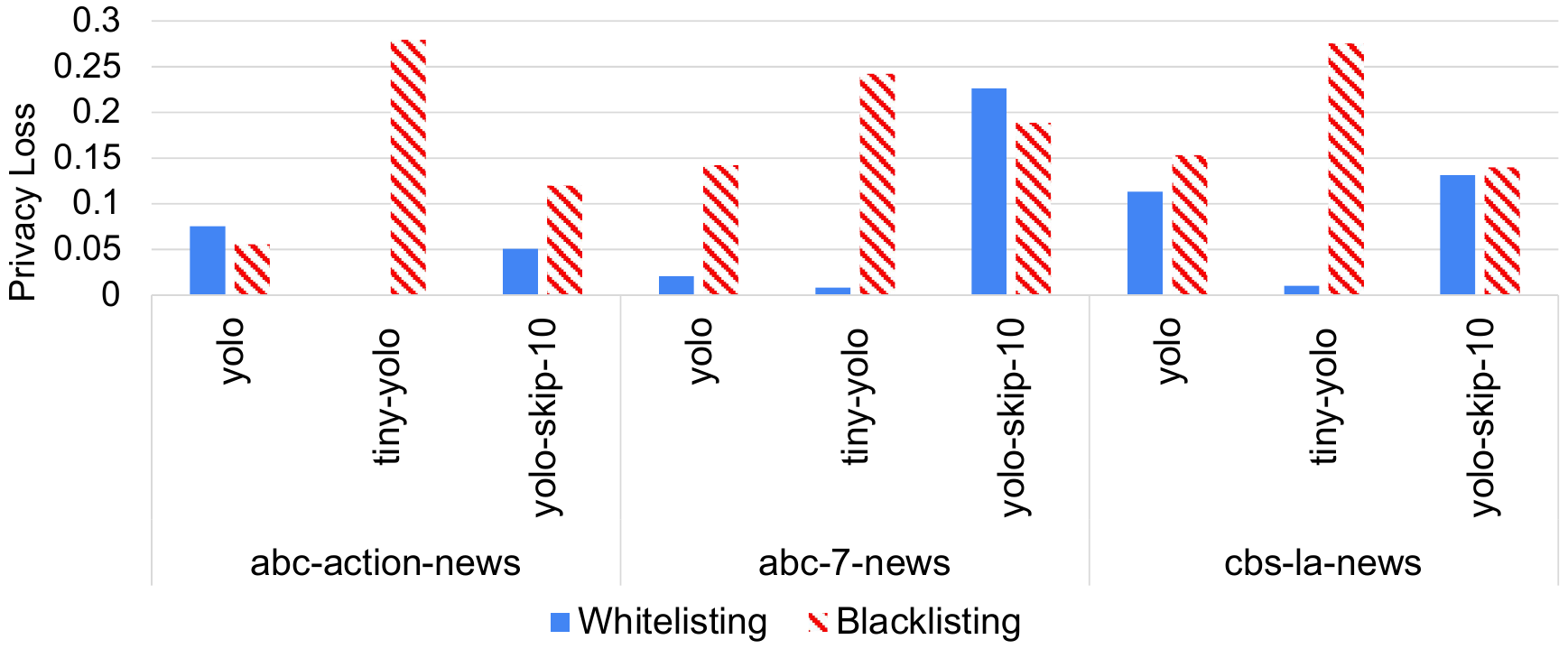} 
	\vspace{-0.6cm}
	\caption{Fraction of faces detected by the face-detection app, comparing whitelisting (\textsf{car}, \textsf{truck}, \textsf{bus} disclosed) and blacklisting (\textsf{person} redacted)}
	\label{fig:face-detection-privacy}
\end{figure}

\textbf{Privacy Experiments:} To measure privacy loss we consider the face-detection app as an adversary and run it on the whitelisted and blacklisted streams that we created for the vehicle-counting application, where: (i) the whitelisted stream is only supposed to disclose objects belonging to the classes \textsf{car}, \textsf{truck} and \textsf{bus}, and (ii) the blacklisted stream redacts all objects belonging to class \textsf{person}. In an ideal world, the face-detection app should not be able to see a face of \textit{any} \textsf{person} in these streams. However, 
some faces may show up due to detection/mis-detection errors. Therefore, we measure the fraction of faces detected in these \textit{privacy-preserving} video streams, as compared to the raw video.

Figure \ref{fig:face-detection-privacy} plots the fraction of faces (privacy loss) detected by the face-detection app, comparing whitelisting (\textsf{car}, \textsf{truck} and \textsf{bus} disclosed) and blacklisting (\textsf{person} redacted), for different object-detection techniques. In most cases we observe that blacklisting can lead to higher privacy leakage than whitelisting -- up to \textasciitilde28x higher. However, for video \textsf{abc-7-news} using \textsf{yolo-skip-10}, we observe that it is whitelisting that leads to a slightly higher privacy loss (\textasciitilde1.2x). 

We can conclude that, while \system preserves privacy, its composable streams can also be easily utilized by unmodified applications, with minimal utility loss as compared to using an un-redacted video. Additionally, the availability of per-object class metadata streams can enable many commonly found simple object-counting applications.

\section{Related Work} \label{sec:related-work}
We now briefly describe the relevant prior work.

\textbf{Object Detection:} CNNs \cite{cnn} are the basis of most state-of-the-art techniques for object-detection \cite{yolo, faster-rcnn, ssd}. 
A more complex and computationally-intensive task is that of image segmentation \cite{image-segmentation-1, image-segmentation-2}, which involves assigning each pixel in the image a class label. \system can also use segmentation to create \textit{more} accurate whitelisted/blacklisted streams with lower privacy leakage. However, segmentation is resource-intensive and not suited for execution on edge platforms.
 
Fast inference is key to live analytics. Therefore, prior work has looked at techniques to speed up inference including: (i) neural-network specialization \cite{focus, nn-specialization-1, nn-specialization-2}, (ii) approximation \cite{nn-approx-1, nn-approx-2}, and (iii) cascading a series of classifiers \cite{cascade-1, cascade-2, cascade-3, cascade-4}. Recent work \cite{noscope} has also looked at exploiting the presence of the same objects across multiple frames to speed up inference on video streams. All these techniques are complementary to \system, and can be plugged in to generate privacy-preserving video streams.

\textbf{Video Analytics:} Recent video-analytics systems \cite{live-analytics, focus, video-edge, chameleon, wide-area-analytics, noscope} have focused on multiple aspects including, indexing video at ingest time to reduce query latency \cite{focus}, scheduling query execution across clusters \cite{video-edge, wide-area-analytics} increasing accuracy \cite{chameleon}, and video/query-specific detector specialization \cite{noscope}. \system is complementary to these systems and provides privacy-preserving live streaming and analytics.

\textbf{Visual Privacy:} Multiple prior works utilize the \textit{blacklisting} approach to redact or distort sensitive information from videos in real time based on: (i) specification \cite{privacy-preserving-satya}, (ii) detected context \cite{life-log, i-pic}, and (iii) visual markers \cite{respectful-cameras}. Video-sharing platforms like YouTube also provide tools for creators to automatically blur out sensitive objects \cite{youtube}. On the other hand, some prior work \cite{recognizer, surround-web, private-eye} has also used the \textit{whitelisting} approach based on the principle of least privilege, to restrict access to objects in a video stream based on user requirements in the context of augmented reality \cite{recognizer, surround-web} and smartphone applications \cite{private-eye}. Video-conferencing systems like Skype \cite{skype} also provide an option to automatically blur out all the background except the person's face. Our work proves that whitelisting is crucial for scalable, privacy-preserving video analytics. Unlike prior work and commercial products which deal with homogeneous consumers, \system also provides a whitelisting-inspired bandwidth-efficient scheme to distribute privacy-preserving video streams to multiple consumers with different requirements. Our scheme for encrypting each object-stream with a different key builds on the \textsf{P3} photo-sharing mechanism \cite{p3}, which encrypts all \textit{sensitive} information in a photo with a single key. 

Recent work has also looked at obfuscating video streams to reduce privacy leakage \cite{olympus,vgan,game-theory}. These approaches require significant effort to train an obfuscation mechanism, and also require applications to be modified or re-trained to work with the obfuscated video. This obfuscated video may also be unsuitable for human viewing. \system works with off-the-shelf object detectors and provides human-viewable video streams that are usable by unmodified applications.

\textbf{Feature Extraction:} In \cite{Ossia2020DeepPE}, the authors propose a collaborative framework for private feature extraction at the edge (smartphone) based on a verifiable feature-extractor module provided by the data consumer (cloud). This framework allows the data producer to control which private features are used by the data consumer, and is useful for performing fine-grained data sharing. In contrast, to be usable by a large set of unmodified applications, \system performs data transformation using an object detector which extracts features at the scale of individual objects. 

\textbf{Object-Detection Attacks:} Recent work has also looked at scenarios where malicious actors can \textit{induce} object-detection errors by masquerading as other entities \cite{mahmood-masquerade, papernot2017practical}. In response, techniques have been proposed to train detectors/classifiers to be robust against such attacks \cite{madry2017towards, tramer2017ensemble}. These techniques can be used to train the object-detection model used by \system, to mitigate the risks of such attacks. 

\textbf{Privacy-Preserving Data Analytics:} Researchers have also looked at techniques based on differential privacy \cite{differential-privacy, pripearl, priv-approx} and secure multi-party computation \cite{du2001secure}. These cryptographic approaches are more suited to structured data and/or require applications to be modified to extract utility.




\section{Conclusion} \label{sec:conclusion}
Machine-learning-based techniques only offer probabilistic guarantees. Therefore, perfect privacy cannot be guaranteed while using object detection to create \textit{whitelisted} or \textit{blacklisted} video streams. This paper has made the case that a whitelisting, or blocking-by-default, approach to redaction is crucial for scalable, privacy-preserving live video analytics. In doing so, we analytically proved that when using modern object-detection techniques at video-ingest time, whitelisting can \textit{guarantee} lower privacy leakage than blacklisting. 

We have designed, implemented, and evaluated a system called \system that embodies this approach. To minimize the size of its trusted computing base, \system uses resource-constrained edge nodes to perform object detection on video streams. For each object category, \system then generates an encrypted, per-category live stream and distributes these encrypted videos to applications. 
Applications retrieve decryption keys for their authorized categories, and can compose decrypted videos to create a coherent multi-object stream. Experiments with our \system prototype show that (i) compared to blacklisting, whitelisting yields significantly better privacy (up to 28x) and bandwidth savings (up to 5.5x), and (ii) \system is usable by \textit{unmodified} real-world applications with negligible utility loss.

While the key ideas proposed in this paper are relevant for video streaming, we believe that they are also applicable for other types of content streaming, including audio and text.

\Urlmuskip=0mu plus 1mu\relax
\bibliographystyle{plain}
\bibliography{sample}

\end{document}